\documentclass[12pt]{article}
\usepackage{epsfig}
\usepackage{psfrag}
\usepackage{latexsym}
\usepackage{indentfirst}
\usepackage{fancyhdr}
\usepackage{amssymb}
\usepackage{amsmath}
\usepackage{amsfonts}
\usepackage{cite}
\usepackage{bbold}
\usepackage{color}

\usepackage{fancybox}
\usepackage[footnotesize]{caption2}
\usepackage{graphicx}
\usepackage[center,footnotesize,hang]{subfigure}
\usepackage{url}
\usepackage{array}
\renewcommand{\baselinestretch}{1.25}%

\begin{document}
\begin{titlepage}
\vspace*{-1cm}

\vskip 2.5cm
\begin{center}
{\large\bf  Bilarge Neutrino Mixing and Abelian Flavor Symmetry}
\end{center}
\vskip 0.2  cm
\vskip 0.5  cm
\begin{center}
{Gui-Jun Ding}$^{a}$,
{S. Morisi}$^{{b},{c}}$, {and J.~W.~F.~Valle}$^{b}$
\\
\vskip .2cm
$^{a}${\it Department of Modern Physics,}
\\
{\it University of Science and Technology of China, Hefei, Anhui
230026, China}
 \vskip .3cm
$^b$ {\it AHEP Group, Instituto de F\'{\i}sica Corpuscular --
    C.S.I.C./Universitat de Val{\`e}ncia \\
    Edificio de Institutos de Paterna, Apartado 22085,
  E--46071 Val{\`e}ncia, Spain}  \vskip.3cm
 \vskip .3cm
$^c$ {\it
Institut f{\"u}r Theoretische Physik und Astrophysik, \\ Universit{\"a}t
W{\"u}rzburg, 97074 W{\"u}rzburg, Germany}
\end{center}
\vskip 0.7cm
\begin{abstract}
\noindent

We explore two bilarge Neutrino Mixing anz$\ddot{a}$tze within the
context of Abelian flavor symmetry theories: ($\mathrm{\tt BL_1}$)
$\sin\theta_{12}\sim\lambda$, $\sin\theta_{13}\sim\lambda$,
$\sin\theta_{23}\sim\lambda$, and ($\mathrm{\tt BL_2}$)
$\sin\theta_{12}\sim\lambda$, $\sin\theta_{13}\sim\lambda$,
$\sin\theta_{23}\sim 1-\lambda$. The first pattern is proposed by two
of us and is favored if the atmospheric mixing angle $\theta_{23}$
lies in the first octant, while the second one is preferred for the
second octant of $\theta_{23}$.
In order to reproduce the second texture, we find that the flavor
symmetry should be $U(1)\times Z_m$, while for the first pattern
the flavor symmetry should be extended to $U(1)\times Z_m\times Z_n$
with $m$ and $n$ of different parity.
Explicit models for both mixing patterns are constructed based on the
flavor symmetries $U(1)\times Z_3\times Z_4$ and $U(1)\times Z_2$. The
models are extended to the quark sector within the framework of
$SU(5)$ grand unified theory in order to give a successful description
of quark and lepton masses and mixing simultaneously. Phenomenological
implications are discussed.

\end{abstract}
\end{titlepage}
\setcounter{footnote}{0}
\vskip2truecm

\section{Introduction}

Our knowledge of the neutrino oscillation parameters has enormously
improved in recent years. In particular the Daya Bay \cite{An:2012eh},
RENO \cite{Ahn:2012nd} and Double Chooz \cite{Abe:2011fz}
Collaborations have established that the reactor mixing angle
$\theta_{13}>0$ at about $5\sigma$ confidence level, confirming the early
hints for a nonzero
$\theta_{13}$~\cite{Abe:2011sj,Adamson:2011qu}. Recent global analyses
\cite{Tortola:2012te,Fogli:2012ua} of neutrino oscillation parameters,
including the data released at the Neutrino-2012 conference, find that
$\theta_{13}$ is non-zero at about $10\sigma$, and non-maximal
atmospheric mixing angle $\theta_{23}$ is preferred. However, it still
isn't clear which octant $\theta_{23}$ lies in. The global fit of
Ref. \cite{Tortola:2012te}, prefers $\theta_{23}$ in the second octant
with the best fit value
$\sin^2\theta_{23}=0.613\,(0.600)$ 
for normal (inverted) neutrino mass hierarchy, although this hint is
quite marginal and first octant values of $\theta_{23}$ are
well inside the $1\sigma$ range for normal hierarchy and at
$1.2\sigma$ for inverted spectrum. While the independent
phenomenological analyses of atmospheric neutrino data in
Ref. \cite{Fogli:2012ua} obtain a preference for $\theta_{23}$ in
the first octant for both mass hierarchies and exclude maximal mixing
at the $2\sigma$ level, the best fit value is found to be
$\sin^2\theta_{23}=0.386\,(0.392)$ for normal (inverted) neutrino
spectrum. Alternative recent global fits claim both the first and
second $\theta_{23}$ octants are possible \cite{GMSS}. As for the
mass-squared difference, the best fit values of $\Delta
m^2_{\mathrm{sol}}$ and $\Delta m^2_{\mathrm{atm}}$ are $7.62\times
10^{-5}\mathrm{eV}^2$ and $2.55(2.43)\times 10^{-3}\mathrm{eV}^2$
respectively, which lead to $\Delta m^2_{\mathrm{sol}}/\Delta
m^2_{\mathrm{atm}}\simeq0.030(0.031)$. Here the values shown in
parentheses correspond to the inverted neutrino mass hierarchy. Note
that the three groups give almost the same $3\sigma$ ranges for the
lepton mixing parameters.

From the theoretical or model-building point of view, one implication
of this significant experimental progress is that it excludes the
tri-bimaximal mixing ansatz for neutrino mixing \cite{TBM}, unless the
underlying theory is capable of providing sufficiently large
corrections. So far many suggestions have been advanced to explain the
new data, in particular the largish $\theta_{13}$
\cite{theta_13,Toorop:2011jn,Ding:2012xx,King:2012in,Boucenna:2012xb,King:2012vj}.
Instead of seeking for new mass-independent lepton mixing matrices to
replace the tri-bimaximal pattern~\cite{Toorop:2011jn,Ding:2012xx,King:2012in}, which may be
derived from certain discrete flavor symmetries,
Ref.~\cite{Boucenna:2012xb} proposed a novel Wolfenstein-like
\textit{ansatz} for the neutrino mixing matrix. In this scheme, all
three lepton mixing angles are assumed to be of the same order to
first approximation
\begin{equation}
\label{eq:BL1}\sin\theta_{12}\sim\lambda,\quad\quad\sin\theta_{13}\sim\lambda,\quad\quad \sin\theta_{23}\sim\lambda,
\end{equation}
where $\lambda\simeq0.23$ is the Cabibbo angle, and the symbol
``$\sim$" implies that the above relations contain unknown factors of
order one, the freedom in these factors can be used to obtain an
adequate description of the neutrino mixing. Inspecting the global
data fitting \cite{Tortola:2012te,Fogli:2012ua,GMSS}, we see that
$\sin\theta_{12}\simeq2.5\lambda$ and
$\sin\theta_{13}\simeq\lambda/\sqrt{2}$, which is proposed in the
so-called Tri-bimaximal-Cabibbo mixing \cite{King:2012vj} and also
appeared in the context of quark-lepton complementarity
\cite{Minakata:2004xt}. Such bilarge mixing
pattern~\cite{Boucenna:2012xb} would clearly provide a good leading
order approximation for the current neutrino mixing pattern, if the
atmospheric neutrino mixing angle $\theta_{23}$ turns out to lie in
the first octant. However, the second octant of $\theta_{23}$ can not
be ruled out and is supported by the analyses in
Refs.~\cite{Tortola:2012te,GMSS}.  In this case the texture
\begin{equation}
\label{eq:BL2}\sin\theta_{12}\sim\lambda,\quad\quad\sin\theta_{13}\sim\lambda,\quad\quad \sin\theta_{23}\sim1-\lambda
\end{equation}
could be taken as a viable model-building standard. We shall refer to
two mixing patterns as $\mathrm{\tt BL_1}$ and $\mathrm{\tt BL_2}$
textures respectively. The difference between $\mathrm{\tt BL_1}$ and
$\mathrm{\tt BL_2}$ mixing lies in the order of magnitude of the
atmospheric mixing angle $\theta_{23}$; the $\mathrm{\tt BL_1}$ mixing
pattern would be favored if future experiments establish that
$\theta_{23}$ belongs to the first octant and the deviation from
maximal mixing is somewhat large; otherwise, $\mathrm{\tt BL_2}$ mixing
is preferred. It is well-known that the observed hierarchies of masses
and flavor mixing in the quarks and charged leptons sectors can be
conveniently characterized by the Cabibbo angle. As a result the
$\mathrm{\tt BL_1}$ and $\mathrm{\tt BL_2}$ parametrization may have
deep implications for the theoretical formulation of the ultimate
unified theory of flavor. A lot of work in the literature has
demonstrated that the smallness and hierarchy of the quark masses and
mixing angles can be naturally generated in theories which, at low
energy, are described effectively by an Abelian horizontal symmetry,
which is explicitly broken by a small parameter
\cite{FN,Leurer:1992wg,Dudas:1996fe}. It certainly follows a natural
path to try and apply these ideas on Abelian family symmetries
developed for the quarks to the lepton sector.
In this work, we shall investigate whether and how the $\mathrm{\tt
  BL_1}$ and $\mathrm{\tt BL_2}$ textures can be reproduced naturally
from the Abelian horizontal flavor symmetry. For generality we assume
that the light neutrino masses arise from lepton-number-violating
effective Weinberg-like operators.

The paper is organized as follows. In section \ref{sec:th}, we present
the effective low energy theory for the Abelian $U(1)$ flavor symmetry
and its extension to $U(1)\times Z_m\times Z_n$. We find that, in
order to produce the $\mathrm{\tt BL_1}$ texture without fine-tuning,
the family symmetry should be $U(1)\times Z_m\times Z_n$ with $m$ and
$n$ of opposite parity. Models for the $\mathrm{\tt BL_1}$ and
$\mathrm{\tt BL_2}$ schemes are constructed in section \ref{sec:BL1}
and section \ref{sec:BL2} respectively. These models are extended
to include quarks within the $SU(5)$ grand unified theory (GUT), the
observed patterns of both quark and lepton masses and flavor mixings
are reproduced, and the general phenomenological predictions of the
models are discussed. Finally, our conclusions are summarized in
section \ref{sec:conclusion}.


\section{\label{sec:th}Theoretical Framework}

Our theoretical framework is defined as follows. For definiteness we
consider a low energy effective theory with the same particle content
as the supersymmetric Standard Model (SM). In addition to
supersymmetry and the SM gauge symmetry, we introduce a horizontal
$U(1)$ symmetry and a SM singlet chiral superfield $\Theta$ which is
charged under the $U(1)$ family symmetry; without loss of generality,
we normalize its charge to $-1$. The effective Yukawa couplings of the
quarks and leptons are generated from nonrenormalizable superpotential
terms of the form
\begin{eqnarray}
\nonumber&&W=(y_u)_{ij}Q_iU^c_jH_u\left(\frac{\Theta}{\Lambda}\right)^{F(Q_i)+F(U^c_j)}+(y_d)_{ij}Q_iD^c_jH_d\left(\frac{\Theta}{\Lambda}\right)^{F(Q_i)+F(D^c_j)}\\
\label{eq:U(1)}&&\quad+(y_e)_{ij}L_iE^c_jH_d\left(\frac{\Theta}{\Lambda}\right)^{F(L_i)+F(E^c_j)}+(y_{\nu})_{ij}\frac{1}{\Lambda}L_{i}L_{j}H_{u}H_{u}\left(\frac{\Theta}{\Lambda}\right)^{F(L_i)+F(L_j)}
\end{eqnarray}
where $H_{u,d}$ are Higgs doublets, $Q_i$ and $L_i$ are the
left-handed quark and lepton doublets respectively, $U^c_j$, $D^{c}_j$
and $E^c_j$ are the right-handed up-type quark, down-type quark and
charged lepton superfields respectively, and $i$, $j$ are generation
indices. The parameter $\Lambda$ is the cutoff scale of the $U(1)$
symmetry, and $F(\psi)$ denotes the $U(1)$ charge of the field
$\psi$. Note that $F(H_u)$ and $F(H_d)$ do not appear in the
exponents since one can always set the horizontal charges of the
Higgs doublet $H_u$ and $H_u$ to zero by redefinition of the $U(1)$
charges. The last term of Eq.(\ref{eq:U(1)}) is the high-dimensional
version of the effective lepton-number-violating Weinberg operator.

For the Froggatt-Nielsen flavon field $\Theta$, the supersymmetric action contains a Fayet-Iliopoulos term and the associated D-term in the scalar potential provides a large vacuum expectation value (VEV) for the scalar component of $\Theta$. The D-term in the potential is given by
\begin{equation}
V_D=\frac{1}{2}(M^2_{FI}-g_{\Theta}|\Theta|^2)^2
\end{equation}
where $M^2_{FI}$ is the Fayet-Iliopoulos term. The vanishing of $V_D$ requires
\begin{equation}
|\langle\Theta\rangle|=M_{FI}/\sqrt{g_{\Theta}}
\end{equation}
We note that this flavor symmetry breaking mechanism is also
frequently exploited in discrete flavor symmetry model building
\cite{Altarelli:2009gn}. Once the horizontal symmetry is broken by
the VEV $\langle\Theta\rangle$, one obtains the quark and lepton mass
matrices whose elements are suppressed by powers of the small
parameter $\langle\Theta\rangle/\Lambda$, which for simplicity is
usually assumed to be characterized by the Cabibbo angle, i.e.,
$\lambda=\langle\Theta\rangle/\Lambda$, then we have
\begin{eqnarray}
\nonumber&&(M_u)_{ij}=(y_u)_{ij}\lambda^{F(Q_i)+F(U^c_j)}v_u,\quad\quad (M_d)_{ij}=(y_d)_{ij}\lambda^{F(Q_i)+F(D^c_j)}v_d, \\
\label{eq:mass}&&(M_{e})_{ij}=(y_e)_{ij}\lambda^{F(L_i)+F(E^c_j)}v_u,\quad\quad  (M_{\nu})_{ij}=(y_{\nu})_{ij}\lambda^{F(L_i)+F(L_j)}\frac{v^2_u}{\Lambda}\,
\end{eqnarray}
where $v_{u,d}=\langle H_{u,d}\rangle$ is the electroweak scale VEV of
the Higgs doublet $H_{u,d}$. The factors $(y_u)_{ij}$, $(y_d)_{ij}$,
$(y_e)_{ij}$ and $(y_{\nu})_{ij}$ are not constrained by the flavor
symmetry and are usually assumed to be of order one, the freedom in
these factors is used in order to obtain a quantitative description of
the fermion masses and flavor mixings. Since the holomorphicity of the
superpotential forbids nonrenormalizable terms with a negative power
of the superfield $\Theta$, one has $(M_u)_{ij}=0$ if
$F(Q_i)+F(U^c_j)<0$. Similarly $(M_d)_{ij}=0$ if $F(Q_i)+F(D^c_j)<0$,
$(M_{e})_{ij}=0$ if $F(L_i)+F(E^c_j)<0$, and $(M_{\nu})_{ij}=0$ if
$F(L_i)+F(L_j)<0$.

In our framework, the light neutrino masses are generated by the high-dimensional effective Weinberg operators shown in the last term of
Eq.(\ref{eq:U(1)}), consequently, the light neutrinos are Majorana
particles and its mass matrix $M_{\nu}$ is symmetric with
$(M_{\nu})_{ij}=(M_{\nu})_{ji}$~\footnote{ Note that if we introduce
three right-handed neutrino superfields $N^c_i$ to generate light
neutrino mass via type I seesaw mechanism, the structure of the
light neutrino mass matrix $M_{\nu}$ is independent of the $N^c_i$
charge assignments \cite{Rasin:1993kj,Grossman:1995hk}, unless there
are holomorphic zeros in neutrino Dirac mass matrix $M_D$ or in
Majorana mass matrix $M_N$ for the heavy fields $N^c$.}.
Furthermore, if all the horizontal charges are positive, the
hierarchial structure of the mass matrices shown in Eq.(\ref{eq:mass})
allows a simple order of magnitude estimate for the various mass
ratios and mixing angles:
\begin{eqnarray}
\nonumber&&\frac{m_{u_i}}{m_{u_j}}\sim\lambda^{F(Q_i)-F(Q_j)+F(U^c_i)-F(U^c_j)},\quad \frac{m_{d_i}}{m_{d_j}}\sim\lambda^{F(Q_i)-F(Q_j)+F(D^c_i)-F(D^c_j)}, \quad V_{ij}\sim\lambda^{F(Q_i)-F(Q_j)}, \\
\label{eq:estimate}&&\frac{m_{i}}{m_{j}}\sim\lambda^{2[F(L_i)-F(L_j)]},\quad\quad \frac{m_{\ell_i}}{m_{\ell_j}}\sim\lambda^{F(L_i)-F(L_j)+F(E^c_i)-F(E^c_j)},\quad\quad \sin\theta_{ij}\sim\lambda^{F(L_i)-F(L_j)}\,
\end{eqnarray}
where $m_i$ is the light neutrino mass, and $V_{ij}$ denotes the element
of the quark Cabibbo-Kobayashi-Maskawa quark-mixing matrix (CKM) mixing matrix. We note that the sign $``\sim"$
implies that there is an unknown order one coefficient in each
relation, so that the actual value of the mass ratios and mixing
angles may slightly depart from the naive ``power counting''
estimate. Moreover, if some fields carry negative $F$ charges, then
holomorphy plays an important role and the estimates
(\ref{eq:estimate}) could be violated as well. For the $\mathrm{\tt
  BL_1}$ mixing pattern, both $\sin\theta_{12}$ and $\sin\theta_{23}$
are of order $\lambda$, then we should require
\begin{equation}
F(L_1)=F(L_2)+1,\qquad  F(L_2)=F(L_3)+1
\end{equation}
This implies $F(L_1)=F(L_3)+2$, as a result, we have
$\sin\theta_{13}\sim\lambda^2$. Therefore we conclude that the
$\mathrm{\tt BL_1}$ mixing pattern can not be naturally produced from
a pure $U(1)$ flavor symmetry. Turning to the $\mathrm{\tt BL_2}$
mixing pattern given by $\sin\theta_{23}\sim1$,
$\sin\theta_{12}\sim\lambda$ and $\sin\theta_{13}\sim\lambda$, one
should choose
\begin{equation}
F(L_2)=F(L_3)=F(L_1)-1
\end{equation}

\vskip10.mm

Then we have the $(2i)$ and $(3i)$ $(i=1,2,3)$ entries of the charged
lepton mass matrix are of the same order, hence the diagonalization of
the charged lepton mass matrix leads to large 2-3 mixing. In addition,
we obtain
\begin{equation}
M_{\nu}\sim\lambda^{2F(L_3)}\left(\begin{array}{ccc}
\lambda^2     &   \lambda      &    \lambda   \\
\lambda       &     1          &     1   \\
\lambda       &     1          &     1
\end{array}\right)\frac{v^2_u}{\Lambda}
\end{equation}\label{mnuIII}
Clearly the (2-3) sector of the light neutrino mass matrix has a
democratic structure, thus large mixing in this (2-3) sector is naturally
obtained. However, barring the presence of special cancellations, the
masses of the second and the third light neutrinos are typically
expected to be of the same order in this case. As a result, the three
light neutrinos are quasi-degenerate and strong parameter fine-tuning
is required in order to account for the hierarchy between the measured
mass squared differences $\Delta m^2_{\mathrm{sol}}$ and $\Delta
m^2_{\mathrm{atm}}$.

\vskip10.mm

In order to avoid this kind of fine-tuning in obtaining an acceptable
pattern of neutrino oscillation parameters, we must go beyond the pure
$U(1)$ flavor symmetry case considered above. Let us now move to the
extended flavor symmetry $U(1)\times Z_m\times Z_n\subset U(1)\times U(1)'\times U(1)''$. This kind of
Abelian symmetry is somewhat complex and not yet fully discussed, as
far as we know, since most of the previous work concentrated on $U(1)$ or
$U(1)\times Z_m\subset U(1)\times U(1)'$ flavor symmetry.
We now consider~\cite{Leurer:1992wg,Grossman:1995hk} three SM singlet
superfields $\Theta_1$, $\Theta_2$ and $\Theta_3$ with the horizontal
charges
\begin{equation}
\Theta_1~:(-1,0,0),\quad\quad  \Theta_2~:(0,-1,0),\quad\quad  \Theta_3~:(0,0,-1)
\end{equation}
In exactly the same way as the single $U(1)$ case, the three flavons $\Theta_1$, $\Theta_2$ and $\Theta_3$ could get non-vanishing VEVs determined by corresponding the D-terms. In general the VEVs
$\langle\Theta_1\rangle$, $\langle\Theta_2\rangle$ and
$\langle\Theta_3\rangle$ are different \cite{Leurer:1992wg,Grossman:1995hk}.  For simplicity, we take in what follows:
$\langle\Theta_1\rangle/\Lambda\sim\lambda$,
$\langle\Theta_2\rangle/\Lambda\sim\lambda$ and
$\langle\Theta_3\rangle/\Lambda\sim\lambda$.  The effective Yukawa
couplings are given by extending Eq.(\ref{eq:U(1)}) with new flavons
$\Theta_1$, $\Theta_2$ and $\Theta_3$ as follows:
\begin{eqnarray}
\nonumber&& W=(y_u)_{ij}Q_iU^c_jH_u\left(\frac{\Theta_1}{\Lambda}\right)^{F(Q_i)+F(U^c_j)}\left(\frac{\Theta_2}{\Lambda}\right)^{\left[Z_m(Q_i)+Z_m(U^c_j)\right]}\left(\frac{\Theta_3}{\Lambda}\right)^{\left[Z_n(Q_i)+Z_n(U^c_j)\right]}\\
\nonumber&&\qquad +(y_d)_{ij}Q_iD^c_jH_d\left(\frac{\Theta_1}{\Lambda}\right)^{F(Q_i)+F(D^c_j)}\left(\frac{\Theta_2}{\Lambda}\right)^{\left[Z_m(Q_i)+Z_m(D^c_j)\right]}\left(\frac{\Theta_3}{\Lambda}\right)^{\left[Z_n(Q_i)+Z_n(D^c_j)\right]}\\
\nonumber&&\qquad +(y_e)_{ij}L_iE^c_jH_d\left(\frac{\Theta_1}{\Lambda}\right)^{F(L_i)+F(E^c_j)}\left(\frac{\Theta_2}{\Lambda}\right)^{\left[Z_m(L_i)+Z_m(E^c_j)\right]}\left(\frac{\Theta_3}{\Lambda}\right)^{\left[Z_n(L_i)+Z_n(E^c_j)\right]}\\
\label{eq:U(1)_extended}&&\qquad +(y_{\nu})_{ij}\frac{1}{\Lambda}L_{i}L_{j}H_{u}H_{u}\left(\frac{\Theta_1}{\Lambda}\right)^{F(L_i)+F(L_j)}\left(\frac{\Theta_2}{\Lambda}\right)^{\left[Z_m(L_i)+Z_m(L^{}_j)\right]}\left(\frac{\Theta_3}{\Lambda}\right)^{\left[Z_n(L_i)+Z_n(L^{}_j)\right]}
\end{eqnarray}
where $Z_{m,n}(\psi)$ is the $Z_{m,n}$ charge of the field $\psi$, and
the brackets $[\ldots]$ around the exponents denote that we are
modding out by $m$ ($n$) according to the $Z_m$ ($Z_n$) addition rule,
namely,
\begin{equation}\label{pmm}
\left[Z_m(Q_i)+Z_m(U^c_j)\right] =
\left\{
\begin{array}{lll}
r  &\mbox{if}&  r<m\\
r-m  &\mbox{if}&  r\ge m\\
\end{array}\right. 
\end{equation}
where $r=Z_m(Q_i)+Z_m(U^c_j)$.  We note that the charge assignments of
the Higgs doublets $H_u$ and $H_d$ have been set to $(0,0,0)$ by
redefining the flavor symmetry charges of the fields. Thus, the fermion
mass matrix can be expressed in term of the horizontal charges as
\begin{eqnarray}
\nonumber&&(M_u)_{ij}=(y_{u})_{ij}\lambda^{F(Q_i)+F(U^c_j)+\left[Z_m(Q_i)+Z_m(U^c_j)\right]+\left[Z_n(Q_i)+Z_n(U^c_j)\right]}v_u\,, \\
\nonumber&&(M_d)_{ij}=(y_{d})_{ij}\lambda^{F(Q_i)+F(D^c_j)+\left[Z_m(Q_i)+Z_m(D^c_j)\right]+\left[Z_n(Q_i)+Z_n(D^c_j)\right]}v_d\,,  \\
\nonumber&&(M_{e})_{ij}=(y_{e})_{ij}\lambda^{F(L_i)+F(E^c_j)+\left[Z_m(L_i)+Z_m(E^c_j)\right]+\left[Z_n(L_i)+Z_n(E^c_j)\right]}v_d\,,  \\
&&(M_{\nu})_{ij}=(y_{\nu})_{ij}\lambda^{F(L_i)+F(L_j)+\left[Z_m(L_i)+Z_m(L^{}_j)\right]+\left[Z_n(L_i)+Z_n(L^{}_j)\right]}\frac{v^2_u}{\Lambda}\,.
\end{eqnarray}

Consider the quark sector, the flavor mixing angles there are given by
\begin{eqnarray}
  V_{ij}^{u}&\sim&\lambda^{(F(Q_i)+F(U^c_j))-(F(Q_j)+F(U^c_j))+\big[Z_{m(n)}(Q_i)+Z_{m(n)}(U^c_j)\big]-\big[Z_{m(n)}(Q_j)+Z_{m(n)}(U^c_j)\big]}\,,\\
  V_{ij}^{d}&\sim&\lambda^{(F(Q_i)+F(D^c_j))-(F(Q_j)+F(D^c_j))+\big[Z_{m(n)}(Q_i)+Z_{m(n)}(D^c_j)\big]-\big[Z_{m(n)}(Q_j)+Z_{m(n)}(D^c_j)\big]}\,.
\end{eqnarray}
For $m=n=0$ the CKM matrix elements describing the charged current
weak interaction of quarks behave approximatively as $V_{ij}^{u,d}\sim
\lambda^{F(Q_i)-F(Q_j)}$ and therefore the CKM mixing
$V_{CKM}={V^u}^\dagger\cdot V^d$ is expected to scale as
${V_{CKM_{ij}}}\sim \lambda^{F(Q_i)-F(Q_j)}$.
In order to compare with the pure $U(1)$ horizontal symmetry case, we
can define an effective flavor charge in the general case $m\ne n\ne
0$ as
\begin{equation}
F_{eff}(\psi)=F(\psi)+Z_m(\psi)+Z_n(\psi)\,.
\end{equation}
Then it is clear that
\begin{equation}
  V_{CKM_{ij}}\sim\lambda^{F_{eff}(Q_i)-F_{eff}(Q_j)\pm \alpha m\pm \beta n}\,,
\end{equation}
where $\alpha, \beta=0,1$ and we have used Eq.\,(\ref{pmm}) and the fact that
\begin{eqnarray}
&&\big[Z_m(Q_i)+Z_m(U^c_j)\big]-\big[Z_m(Q_j)+Z_m(U^c_j)\big]=Z_m(Q_i)-Z_m(Q_j)\pm
\alpha m \\
&& \big[Z_m(Q_i)+Z_m(D^c_j)\big]-\big[Z_m(Q_j)+Z_m(D^c_j)\big]=Z_m(Q_i)-Z_m(Q_j)\pm
\alpha m
\end{eqnarray}
where $\alpha=0,1$.  The condition for the value $\pm \beta n$
follows similarly. Likewise for the lepton sector, one obtains
\begin{equation}
V_{ij}^l\sim\lambda^{F_{eff}(L_i)-F_{eff}(L_j)\pm \alpha m\pm \beta n}\,.
\end{equation}

Therefore, the masses and mixing angles can be enhanced or suppressed
by $\lambda^{\pm m\pm n}$ relative to the scaling predictions obtained
when the family symmetry is the continuous flavor symmetry
$U(1)\times U(1)' \times U(1)''$ because of the discrete nature of
$Z_m\times Z_n$. Note that in the case where the light neutrino masses
are generated by the type I seesaw mechanism and all fermion charges are
positive, the neutrino masses and mixing angles still do not depend on
the details of the right-handed neutrino sector, except for the
possible enhancement or suppression associated to the $Z_m\times Z_n$
flavor symmetry.

\vskip10.mm

Furthermore, when the flavor symmetry is reduced to $U(1)\times Z_m$
by taking $n=0$, all the above results remain valid. It is remarkable
that we can employ the $U(1)\times Z_m$ flavor symmetry to maintain
the $\mathrm{\tt BL_2}$ mixing while achieving very different neutrino
masses without fine-tuning. We shall restrict our attention to the
case of a $Z_2$ symmetry which is the minimal nontrivial $Z_m$ group
(see, for example, the explicit model construction given in
sec.~\ref{sec:BL2} below).  In this case just the $Z_m$ symmetry can
reproduce a hierarchy in neutrino masses of order $\lambda^2$
consistent with the observed ratio of solar-to-atmospheric splittings.

In contrast, note that since the reactor neutrino mixing is
necessarily of order $\sin\theta_{13}\sim\lambda^{2\pm \alpha m }$ the
$U(1)\times Z_m$ flavor symmetry can not produce the $\mathrm{\tt
  BL_1}$ mixing pattern.  Indeed for such $\mathrm{\tt BL_1}$ texture
one has $\sin\theta_{12}\sim\lambda$ and $\sin\theta_{23}\sim\lambda$,
which is in conflict with the required linear behavior of the reactor
mixing angle $\sin\theta_{13}\sim\lambda$. Note parenthetically that
the $Z_1$ group consists of only the identity element, so the group
$U(1)\times Z_1$ is isomorphic to $U(1)$, and the $Z_1$ charge of
field is 0, hence the flavor symmetry $U(1)\times Z_1$ produces a
wrong scaling behavior $\sin\theta_{13}\sim\lambda^2$.

We now turn to the realistic case of the $U(1)\times Z_m\times Z_n$
flavor symmetry. If both solar and atmospheric neutrino mixing angles
are of order $\lambda$ then the reactor angle would be constrained to
be of order $\sin\theta_{13}\sim\lambda^{2\pm \alpha m \pm \beta
  n}$. As a result, one can have $\sin\theta_{13}\sim\lambda$ if the
parity of $m$ and $n$ is opposite. This is an interesting observation
of the present work. In section \ref{sec:BL1}, a concrete model for
the $\mathrm{\tt BL_1}$ mixing pattern is presented based on the
flavor symmetry $U(1)\times Z_3\times Z_4$.

Since an Abelian flavor symmetry can not predict the exact value of
the $\mathcal{O}(1)$ coefficients in front of each invariant operator,
we must content ourselves with explaining the orders of magnitude of
fermion masses and flavor mixing parameters. To identify the
phenomenologically acceptable mass matrices, we will estimate the
various mass ratios and mixing angles as approximate powers of the
small parameter $\lambda$. The hierarchies in the quark mixing angles
are clearly displayed in Wolfenstein's truncated
form~\cite{Wolfenstein:1983yz} of the parametrization of the CKM
matrix~\cite{Schechter:1980gr}:
\begin{equation}
V_{CKM}=\left(\begin{array}{ccc}
1-\lambda^2/2  &  \lambda           &   A\lambda^3(\rho-i\eta)   \\
-\lambda       & 1-\lambda^2/2      &  A\lambda^2    \\
A\lambda^3(1-\rho-i\eta)     & -A\lambda^2      &1
\end{array}\right)
\end{equation}
where the quantities $A$, $\rho$ and $\eta$ are experimentally determined to be of order one. Therefore the order of magnitude of the three mixing angles is given in terms of the $\lambda$ as
\begin{equation}
|V_{us}|\sim\lambda,\quad\quad  |V_{cb}|\sim\lambda^2,\quad\quad  |V_{ub}|\sim\lambda^3-\lambda^4
\end{equation}
The charged fermion mass ratios at the grand unified theory (GUT)
scale should satisfy~\cite{mass:ratio}
\begin{eqnarray}
\nonumber&& \frac{m_u}{m_c}\sim\lambda^4,\qquad\qquad\quad \frac{m_c}{m_t}\sim\lambda^3-\lambda^4,   \\
\nonumber&& \frac{m_d}{m_s}\sim\lambda^2,\qquad\qquad\quad \frac{m_s}{m_b}\sim\lambda^2,   \\
&& \frac{m_e}{m_{\mu}}\sim\lambda^2-\lambda^3,\quad\quad\,\,\,\, \frac{m_{\mu}}{m_{\tau}}\sim\lambda^2
\end{eqnarray}
as well as
\begin{equation}
\quad\frac{m_b}{m_{\tau}}\sim1, \qquad\qquad\quad \frac{m_b}{m_t}\sim\lambda^3
\end{equation}
for the intrafamily hierarchy. The first identity is the well-known
$b-\tau$ unification relation. For the neutrinos, we required that the
lepton mixing is of $\mathrm{\tt BL_1}$ or $\mathrm{\tt BL_2}$ type
depending on the octant of $\theta_{23}$. For the quark sector, all the explicit models are properly constructed to meet the requirement $m_t/v_u\sim1$ and $m_b/v_d\sim\lambda^3$



\section{\label{sec:BL1}Model for $\mathrm{\tt BL_1}$ mixing }

As has been shown in the previous section, one can reproduce the
$\mathrm{\tt BL_1}$ texture within the framework of $U(1)\times
Z_m\times Z_n$ family symmetry, where $m$ and $n$ should have
different parity. For concreteness, we shall use $m=3$ and $n=4$ for
our model. For such symmetry choice the possible model realization of
the $\mathrm{\tt BL_1}$ texture is not unique. As a concrete example,
here the horizontal charges of the lepton fields are taken to be
\begin{eqnarray}
\nonumber&&L_{1}:~(4,1,3),\quad\quad L_{2}:~(3,2,2),\quad\quad L_{3}:~(1,1,1),\quad\quad  \\
\label{eq:ass1}&& E^c_{1}:~(3,2,2),\quad\quad E^c_{2}:~(1,2,2),\quad\quad E^c_{3}:~(0,0,0)\,.
\end{eqnarray}
One immediately obtains the charged lepton mass matrix
\begin{equation}
\label{eq:cmm}M_{e}\sim\left(\begin{array}{ccc}
\lambda^{8} & \lambda^6   & \lambda^8 \\
\lambda^{7} & \lambda^5   & \lambda^{7} \\
\lambda^7 &  \lambda^5   & \lambda^3
\end{array}\right)v_d
\end{equation}
which yields the mass ratios
\begin{equation}
\frac{m_e}{m_{\mu}}\sim\lambda^3,\qquad  \frac{m_{\mu}}{m_{\tau}}\sim\lambda^2\,,
\end{equation}
that are consistent with the experimental requirements. For the
charged assignments in Eq.(\ref{eq:ass1}), the light neutrino mass
matrix is given by
\begin{eqnarray}
\label{eq:nmm}&&M_{\nu}\sim\left(
\begin{array}{ccc}
\lambda^{12}  &  \lambda^8  & \lambda^7  \\
\lambda^8  &  \lambda^7  & \lambda^7 \\
\lambda^7  &  \lambda^7  & \lambda^6
\end{array}
\right)\frac{v^2_u}{\Lambda}\,.
\end{eqnarray}
It predicts the light neutrino mass eigenvalues as follows:
\begin{equation}
\label{eq:mass_BL1}m_{1}\sim\lambda^8\,\frac{v^2_u}{\Lambda},\qquad m_{2}\sim\lambda^7\,\frac{v^2_u}{\Lambda},\qquad m_{3}\sim\lambda^6\,\frac{v^2_u}{\Lambda}
\end{equation}
The neutrino mass spectrum is normal hierarchy, this is confirmed by subsequent numerical analysis. It is remarkable that
this model gives rise to $m_{2}/m_{3}\sim\lambda$ and $\Delta
m^2_{\mathrm{sol}}/\Delta m^2_{\mathrm{atm}}\sim\lambda^2$, which is
in excellent agreement with the experimental data. In conventional
$U(1)$ or $U(1)\times Z_m$ flavor symmetries, if any ratio between
neutrino masses is an odd power of the small breaking parameter,
generally the mixing angle between the two neutrinos will vanish
\cite{Grossman:1995hk}. The crucial point is that the element
$(M_{\nu})_{22}$, which would have been $\mathcal{O}(\lambda^{14})$
under the continuous $U(1)\times U(1)' \times U(1)''$ symmetry, is
enhanced to $\mathcal{O}(\lambda^7)$ due to the discrete symmetry
$Z_3\times Z_4$. Diagonalizing the mass matrices in Eq.(\ref{eq:cmm})
and Eq.(\ref{eq:nmm}) by the standard perturbative techniques
described in Refs. \cite{Leurer:1992wg,Grossman:1995hk,Hall:1993ni},
we get the three lepton flavor mixing angles
\begin{equation}
\label{eq:mixing_BL1}\sin\theta_{12}\sim\lambda,\quad\quad  \sin\theta_{13}\sim\lambda,\quad\quad  \sin\theta_{23}\sim\lambda\,. \quad\quad
\end{equation}
Hence the $\mathrm{\tt BL_1}$ pattern is produced automatically. Note
that the solar neutrino mixing $\sin\theta_{12}$ arises from order
$\lambda$ contributions from the diagonalization of both $M_e$ and
$M_{\nu}$, while at leading order the reactor and the atmospheric
neutrino mixing angles receive contribution only from the neutrino
mass matrix $M_{\nu}$. The off-diagonal elements $(M_{\nu})_{13}$ and
$(M_{\nu})_{23}$ are enhanced by $Z_4$ and $Z_3$ respectively, hence
we have $\sin\theta_{13}\sim\lambda$ and $\sin\theta_{23}\sim\lambda$
instead of the naive expectations $\sin\theta_{13}\sim\lambda^5$ and
$\sin\theta_{23}\sim\lambda^4$ characteristic of the continuous flavor
symmetry case.

In the following, we shall extend the model to encompass also quark
sector. Since GUT relates quarks and leptons,
the transformation properties of quark fields can be determined from
those of leptons.
In order to give a successful description of the observed fermion mass
hierarchies and mixings simultaneously under the same flavor symmetry
acting on quarks and leptons we work in the framework of $SU(5)$, for
definiteness. Another motivation of considering $SU(5)$ unification is
the anomaly cancellation. If the $U(1)$ flavor symmetry is gauged then
a general assignment of flavor charges to the fields will be
anomalous. One can imagine the anomaly to be canceled via the
Green-Schwarz mechanism~\cite{Green:1984sg}, however, one must check
whether the correct relations are satisfied~\cite{Ibanez:1994ig}. A
convenient way to ensure that the flavor charges are amenable to
cancellation is to have the flavor symmetry to commute with the
$SU(5)$ group~\cite{Nelson:1997bt}.

Here we propose a model with the quark and lepton matter assignments  manifestly compatible with potential unification within $SU(5)$. A complete study of a realistic grand unified model model addressing the well-known problems such as the doublet-triplet splitting, the proton lifetime and gauge coupling unification, is beyond the scope of the present paper and will be studied elsewhere.

In the conventional $SU(5)$ grand unified theory, the fields $D^c_i$
and $L_i$ of the same generation are assigned to a $\bar{\mathbf{5}}$
multiplet, the fields $Q_i$, $U^c_i$ and $E^c_i$ are unified in the
$\mathbf{10}$ representation. Since the flavor symmetry is required to
commute with the gauge symmetry, this means that the fields in each gauge multiplet transform in the same way under the flavor
symmetry. Consequently, the quantum numbers of the quark fields under
the flavor symmetry $U(1)\times Z_3\times Z_4$ are as follows:
\begin{eqnarray}
\nonumber&& Q_{L1}:~(3,2,2),\quad\quad Q_{L2}:~(1,2,2),\quad\quad Q_{L3}:~(0,0,0)\,, \\
\nonumber&& U^c_{1}:~(3,2,2),\quad\quad\,\,\, U^c_{2}:~(1,2,2),\quad\quad\,\,\, U^c_{3}:~(0,0,0)\,,   \\
\label{eq:qs}&& D^c_{1}:~(4,1,3),\quad\quad\,\,\, D^c_{2}:~(3,2,2),\quad\quad\,\,\, D^c_{3}:~(1,1,1)\,.\quad\quad
\end{eqnarray}
We note that although there are many possible assignments to produce
the $\mathrm{\tt BL_1}$ texture in the neutrino sector, only a few of
them can satisfy the quark sector phenomenological constraints within
$SU(5)$. It is well-known that the minimal $SU(5)$ grand unified
theory predicts that the down-type quark mass matrix is the transpose
of the charged lepton mass matrix, therefore the down-type quarks and
charged lepton masses are closely related : $m_{e}=m_d$,
$m_{\mu}=m_{s}$ and $m_{\tau}=m_{b}$, which are in gross disagreement
with the measured fermion masses and must be corrected
\cite{Georgi:1979df}. This can be done through the contribution of
renormalizable \cite{Georgi:1979df} or non-renormalizable
\cite{Ellis:1979fg} operators to the Yukawa matrices. Following
Ref. \cite{Altarelli:2000fu}, we introduce an additional $U(1)\times
Z_3\times Z_4$ singlet superfield $\Sigma$ transforming as a $\bf{75}$
of $SU(5)$, which has non-renormalizable couplings to fermions of the
form $\bar{\mathbf{5}}\, \mathbf{10}\, H_{\bar{\mathbf{5}}}\, \Sigma
/\Lambda$. The Yukawa couplings of the down-type quark and charged
leptons then arise from the two $SU(5) \times U(1)\times Z_3\times
Z_4$ invariant superpotential terms~\footnote{The $\bf{75}$ could in
principle also give a contribution in the up sector. However,
following Ref.~\cite{Altarelli:2000fu} we neglect such a term since
it is not needed to reproduce the up-type quark masses.}.
\begin{eqnarray}
\nonumber&& W_d= \left(
\mathbf{10}_{i}(C_1)_{ij}\bar{\mathbf{5}}_{j}H_{\bar{\mathbf{5}}} +
\frac{\Sigma}{\Lambda}\, \mathbf{10}_i (C_2)_{ij}\bar{\mathbf{5}}_j
H_{\bar{\mathbf{5}}} \right)
\left(\frac{\Theta_1}{\Lambda}\right)^{F(\mathbf{10}_i)+F(\bar{\mathbf{5}}_j)}
\left(\frac{\Theta_2}{\Lambda}\right)^{\left[Z^{}_3(\mathbf{10}_i)+Z_3(\bar{\mathbf{5}}_j)\right]}\\
\label{eq:dm}&&\qquad\times\left(\frac{\Theta_3}{\Lambda}\right)^{\left[Z^{}_4(\mathbf{10}_i)+Z_4(\bar{\mathbf{5}}_j)\right]} ,
\end{eqnarray}
which, after the scalar components of $\Sigma$ acquires a VEV, lead to:
\begin{eqnarray}
\nonumber&&(\mathbf{Y}_d)_{ij}=\big((C_1)_{ij} + \kappa\, (C_2)_{ij} \big)
\lambda^{F(Q_i)+F(D^c_j)+\left[Z_3(Q_i)+Z_3(D^c_j)\right]+\left[Z_4(Q_i)+Z_4(D^c_j)\right]} \,,\\
\label{eq:um}&&(\mathbf{Y}_e)_{ij}=\big((C_1)_{ij}-3 \kappa\,(C_2)_{ij} \big)\lambda^{F(Q_j)+F(D^c_i)\left[Z_3(Q_j)+Z_3(D^c_i)\right]+\left[Z_4(Q_j)+Z_4(D^c_i)\right]}\,,
\end{eqnarray}
where $\kappa=\langle\Sigma\rangle/\Lambda$, which breaks the
transposition relation between $\mathbf{Y}_d$ and $\mathbf{Y}_e$ and
can explain the difference between down-type quarks and charged lepton
masses. In our numerical fits, we take $\kappa=0.3$ for illustration
and find that realistic values for down-type quarks and charged lepton
masses can be reproduced. The superpotential for the up-type quark
mass is
\begin{equation}
W_u=\mathbf{10}_i(C_3)_{ij}\mathbf{10}_jH_{5}\left(\frac{\Theta_1}{\Lambda}\right)^{F(\mathbf{10}_i)+F(\mathbf{10}_j)}\left(\frac{\Theta_2}{\Lambda}\right)^{\left[Z^{}_3(\mathbf{10}_i)+Z_4(\mathbf{10}_j)\right]}\left(\frac{\Theta_3}{\Lambda}\right)^{\left[Z^{}_4(\mathbf{10}_i)+Z_4(\mathbf{10}_j)\right]}\,,
\end{equation}
where one has $(C_3)_{ij}=(C_3)_{ji}$ due to the constraint of the $SU(5)$ gauge symmetry. Then one can express the effective Yukawa couplings for the up-type quark in terms of the flavor symmetry charges as
\begin{equation}
(\mathbf{Y}_u)_{ij}=(C_3)_{ij}\lambda^{F(Q_i)+F(Q_j)+\left[Z^{}_3(Q_i)+Z_3(Q_j)\right]+\left[Z^{}_4(Q_i)+Z_4(Q_j)\right]}
\end{equation}
With the assignments dictated by Eq.(\ref{eq:qs}), one has the
following patterns for the up- and down-type quark mass matrices,
\begin{eqnarray}
\label{eq:quark_mm}M_{u}\sim\left(
\begin{array}{ccc}
 \lambda ^7 & \lambda ^5 & \lambda ^7 \\
 \lambda ^5 & \lambda ^3 & \lambda ^5 \\
 \lambda ^7 & \lambda ^5 & 1
\end{array}
\right)v_u  , \qquad  M_d\sim\left(
\begin{array}{ccc}
 \lambda ^8 & \lambda ^7 & \lambda ^7 \\
 \lambda ^6 & \lambda ^5 & \lambda ^5 \\
 \lambda ^8 & \lambda ^7 & \lambda ^3
\end{array}
\right)v_d\,,
\end{eqnarray}
which yield
\begin{eqnarray}
|V_{us}|\sim\lambda^2,\quad\quad|V_{cb}|\sim\lambda^2,\quad\quad|V_{ub}|\sim\lambda^4\,. 
\end{eqnarray}
We note that both the up and down quark sector contribute $\lambda^2$
to the mixing element $|V_{us}|$, therefore an accidental enhancement
of $\mathcal{O}(\lambda^{-1})$ among the undetermined order one
coefficients $(C_1)_{ij}$, $(C_2)_{ij}$ and $(C_3)_{ij}$ is required
in order to describe the correct Cabibbo angle. The remaining CKM
mixing angles $|V_{cb}|$ and $|V_{ub}|$ arise solely from the
diagonalization of the down-type quark mass matrix $M_d$. In addition,
the pattern given by Eq.(\ref{eq:quark_mm}) leads to the following
quark mass scalings:
\begin{eqnarray}
\nonumber&& m_u\sim \lambda^7v_u,\quad\quad  m_c\sim \lambda^3v_u,\quad\quad m_t\sim v_u,\quad\quad \\
&&m_d\sim \lambda^8v_d,\quad\quad  m_s\sim \lambda^5v_d,\quad\quad  m_b\sim \lambda^3v_d \,,\quad\quad
\end{eqnarray}
which describe the experimental data satisfactorily.
Note that the second term in Eq.(\ref{eq:dm}) accounts for the mass difference between the down-type quarks and charged leptons, allowing for an acceptable charged fermion mass pattern.

In order to see in a quantitative way how well the model describes the
observed values of the fermion masses and mixings, we perform a
numerical analysis, within three independent different seeding
methods: namely flat, Gaussian and exponential distributions. The modulus of the undetermined order one coefficients are taken to be
random numbers with flat, Gaussian and exponential distributions in
turn, the corresponding phases are varied between 0 and $2\pi$. The
probability density function $f(x)$ of the three distributions is
well known
\begin{equation}
f(x)=\left\{
\begin{array}{cc}
\frac{1}{b-a}   &  a\leq x\leq b \\
0  &       x<a~\mathrm{or}~ x>b
\end{array}
\right.\,.
\end{equation}
For flat distribution, we take $a=1/3$ and $b=3$ for illustration in the present work. In the case of gaussian distribution,
\begin{equation}
f(x)=\frac{a}{\sqrt{2\pi}\sigma}e^{-\frac{(x-\mu)^2}{2\sigma^2}}\,.
\end{equation}
We set the mean $\mu=1$ and the standard deviation $\sigma=1.5$ in our numerical calculation. The probability density function for the exponential distribution
is
\begin{equation}
f(x)=\left\{\begin{array}{cc}
\lambda e^{-\lambda x}  &  x\geq 0  \\
0            &    x<0
\end{array}
\right. \,.
\end{equation}
Its statistic mean is $1/\lambda$, and $\lambda$ is taken to be 1 as a typical value for numerical simulation.
To the extent that our results are independent of the choice of
seeding method, they are robust and not simply an artifact of the
choice of the seed function.

\begin{figure}[hptb!]
\begin{center}
\begin{tabular}{ccc}
\includegraphics[scale=.22]{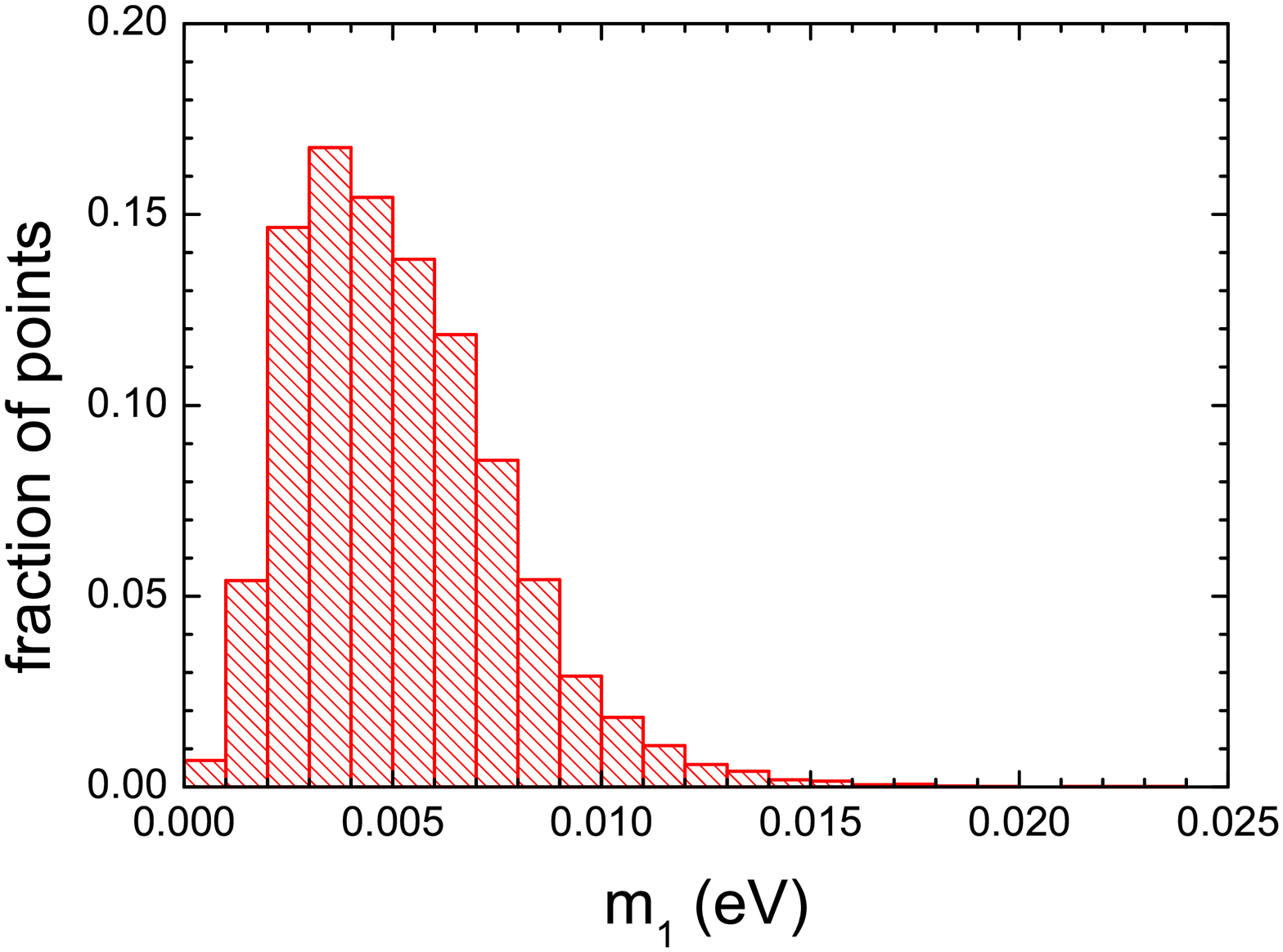}            & \hskip-1cm  \includegraphics[scale=.22]{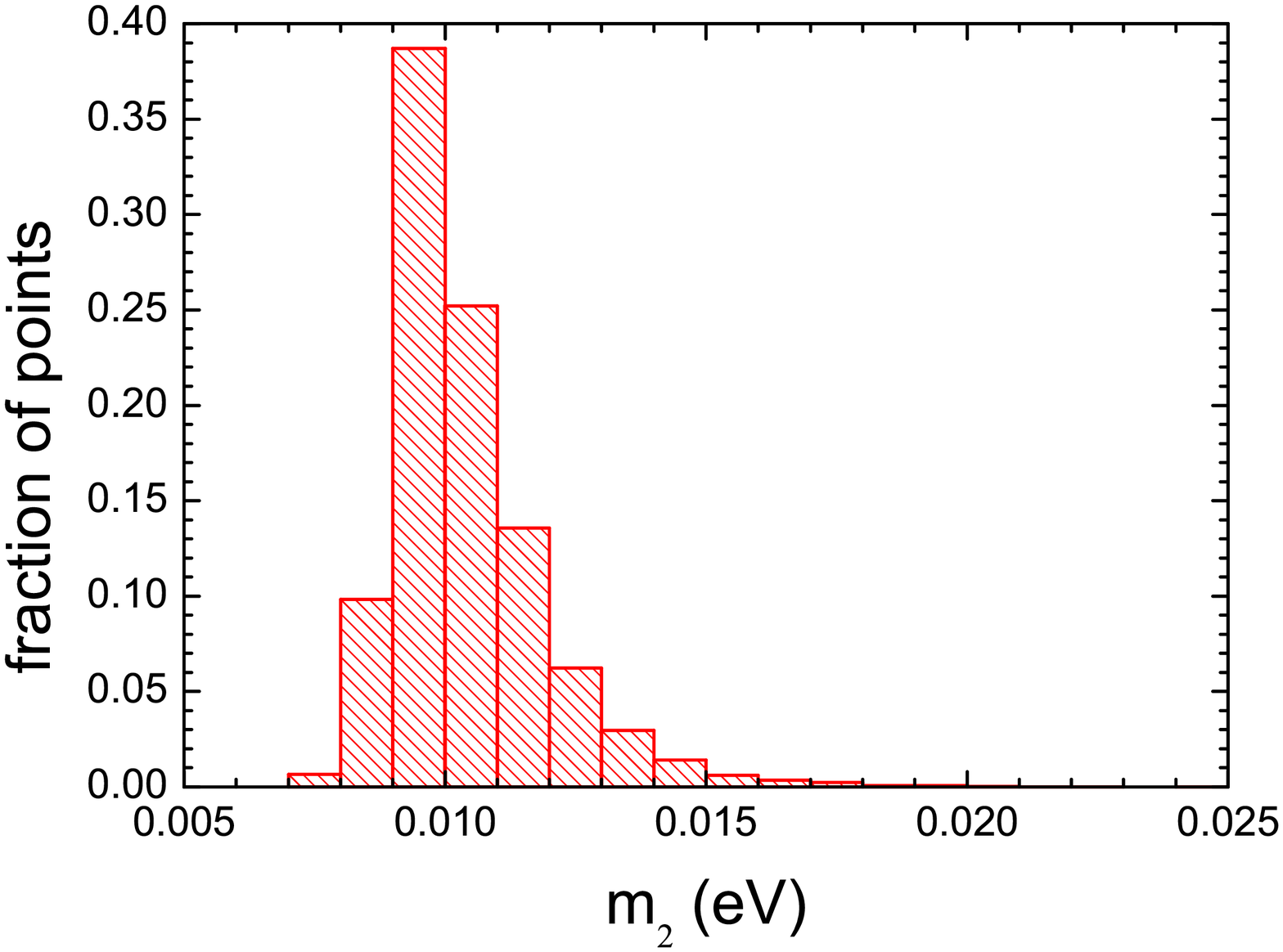}          &  \hskip-1cm   \includegraphics[scale=.22]{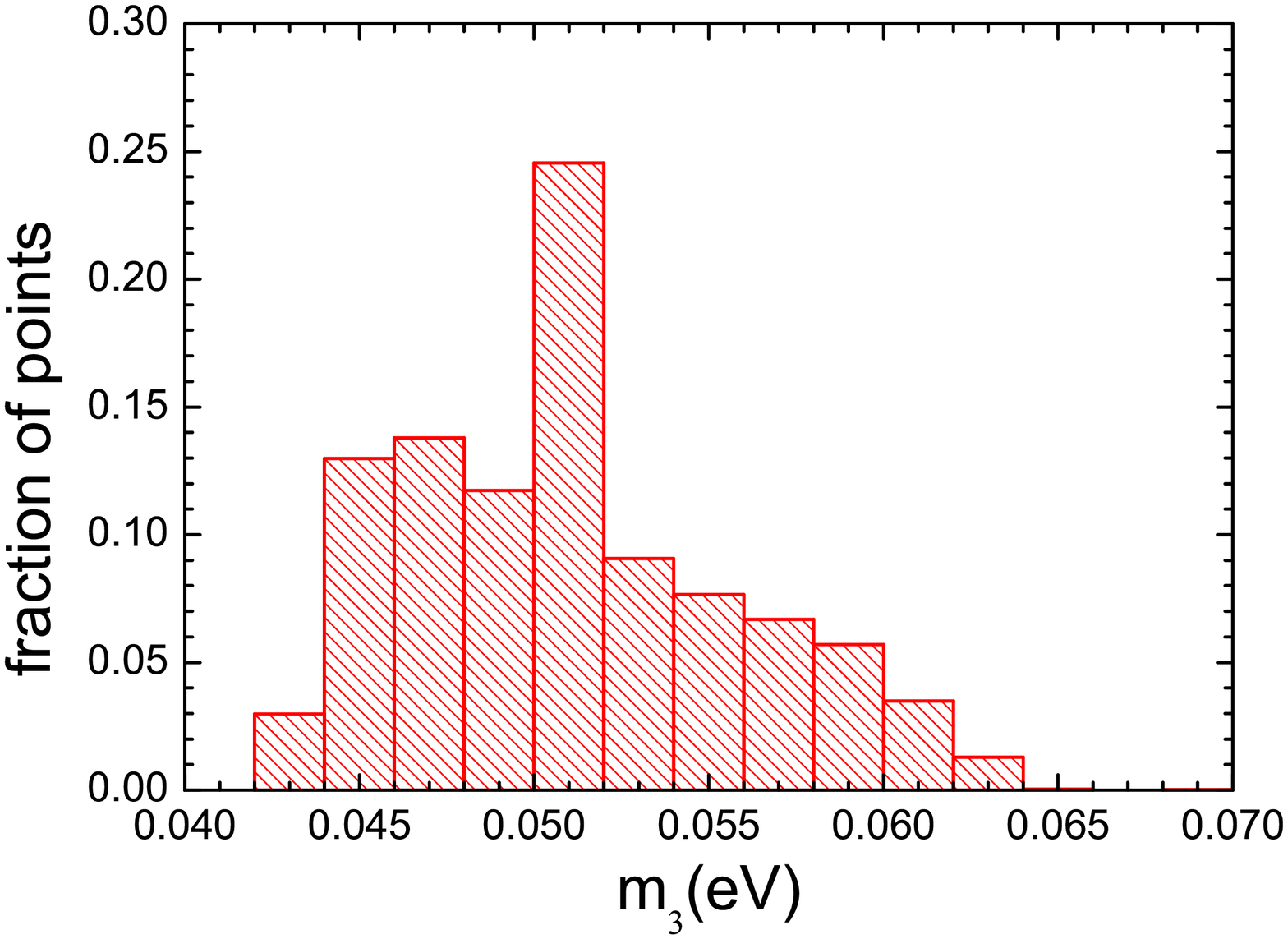} \\
\includegraphics[scale=.20]{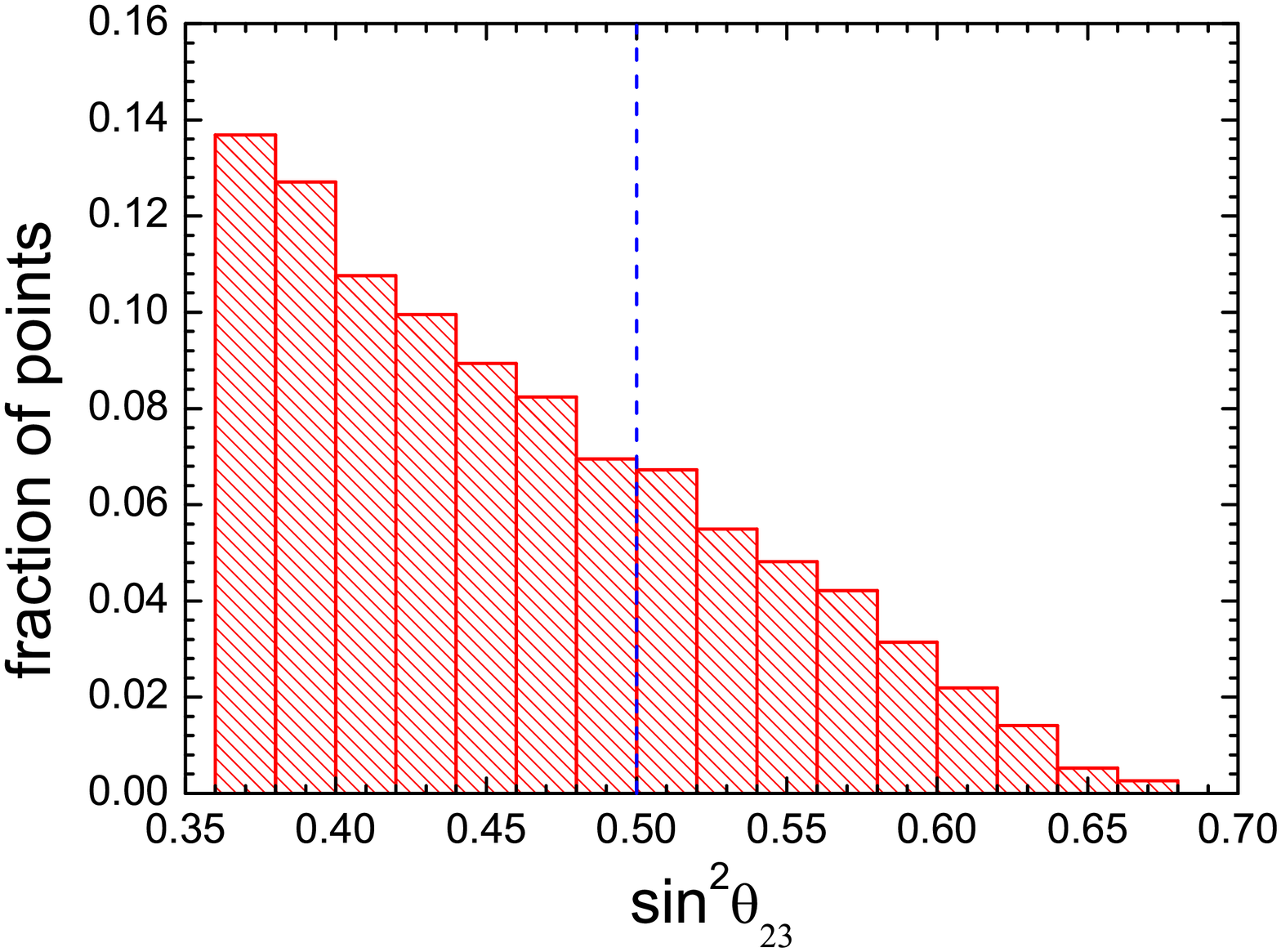}    &   
\hskip-1cm  \includegraphics[scale=.22]{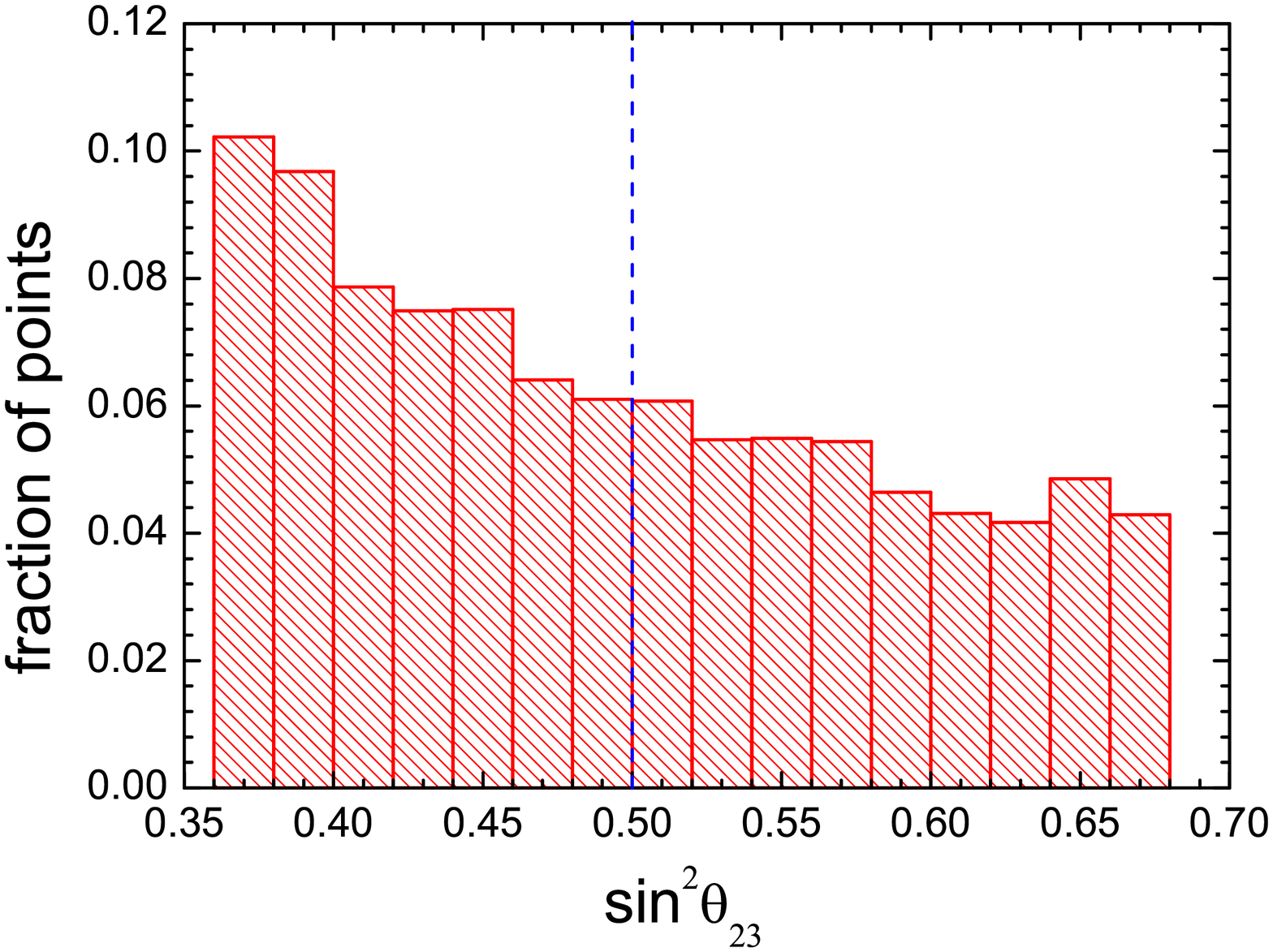}  &  
\hskip-1cm   \includegraphics[scale=.22]{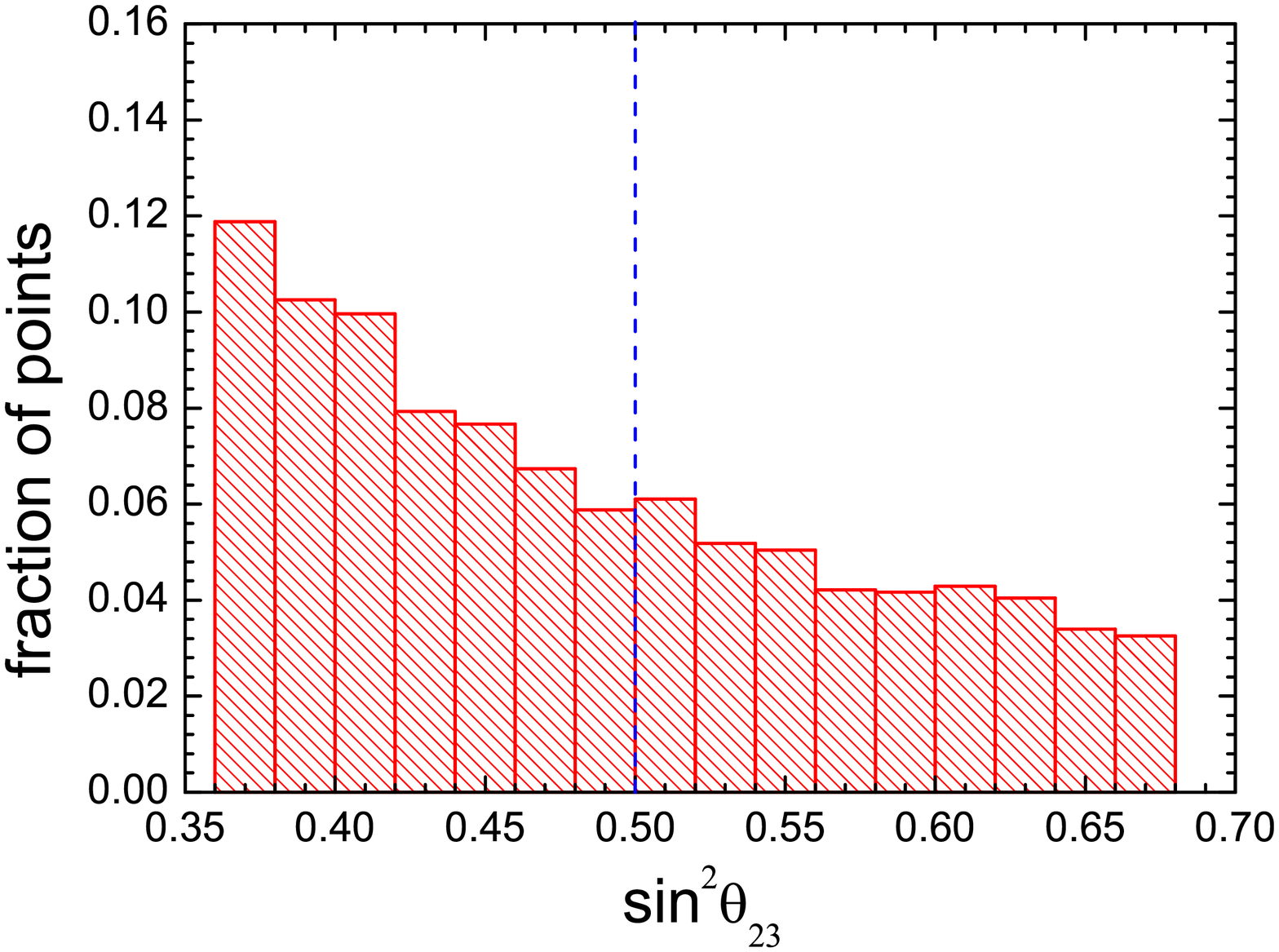}  \\ 
\end{tabular}
\caption{\label{fig:mixing_parameter_BL1} Histograms for the
distribution of light neutrino masses and atmospheric neutrino
mixing parameter in the $\mathrm{\tt BL_1}$ model. In the second row, the left, middle and right panels are obtained with using different seed procedures for the order one Yukawa coefficients, namely flat, exponential and Gaussian, respectively, from left to right.  }
\end{center}
\end{figure}

The coefficients $(C_1)_{ij}$, $(C_2)_{ij}$, $(C_3)_{ij}$ and
$(y_{\nu})_{ij}$ are treated as random complex numbers with arbitrary
phases and absolute value in the interval of $[1/3,3]$. Then we
calculate the quark and lepton masses as well as the CKM and lepton
mixing matrix entries which are required to lie in the experimentally
allowed ranges. The numerical results are found to be nicely
consistent with the above theoretical estimates and qualitative
discussions. Since the flavor parameters of the quark sector are
precisely measured, here we focus on the neutrino sector. As an
example the predicted distributions for the light neutrino masses and
atmospheric mixing parameter are shown in
Fig. \ref{fig:mixing_parameter_BL1}. The light neutrino masses
follow the normal hierarchy pattern and, for all the points produced,
though all non-vanishing, they are rather tiny, with most of the
expected $m_{1}$ values below $0.015\;\mathrm{eV}$. As to the
mixing angles, no specific values of $\theta_{12}$ and $\theta_{13}$
are favored within $3\sigma$, and hence they are not shown in the
figure~\footnote{Similarly, we can hardly see any specific preferred
pattern for the CP violating phases $\delta$, $\varphi_1$ and $\varphi_2$, hence, as before, these are not shown.}.
In contrast, however, the atmospheric neutrino mixing angle
$\theta_{23}$ obeys $\sin^2\theta_{23}<1/2$, which means that
non-maximal $\theta_{23}$ values are preferred, as indicated by
current neutrino oscillation global analyses post-Neutrino
2012~\cite{Tortola:2012te,Fogli:2012ua,GMSS}, with a preference for
the first octant. This has been one of our motivations for introducing
$\mathrm{\tt BL_1}$ mixing pattern, which leads to $\sin\theta_{23}$
of order $\lambda$ at leading order.

\begin{figure}[t!]
\begin{center}
\begin{tabular}{cc}
\hskip-0.5cm \includegraphics[scale=.30]{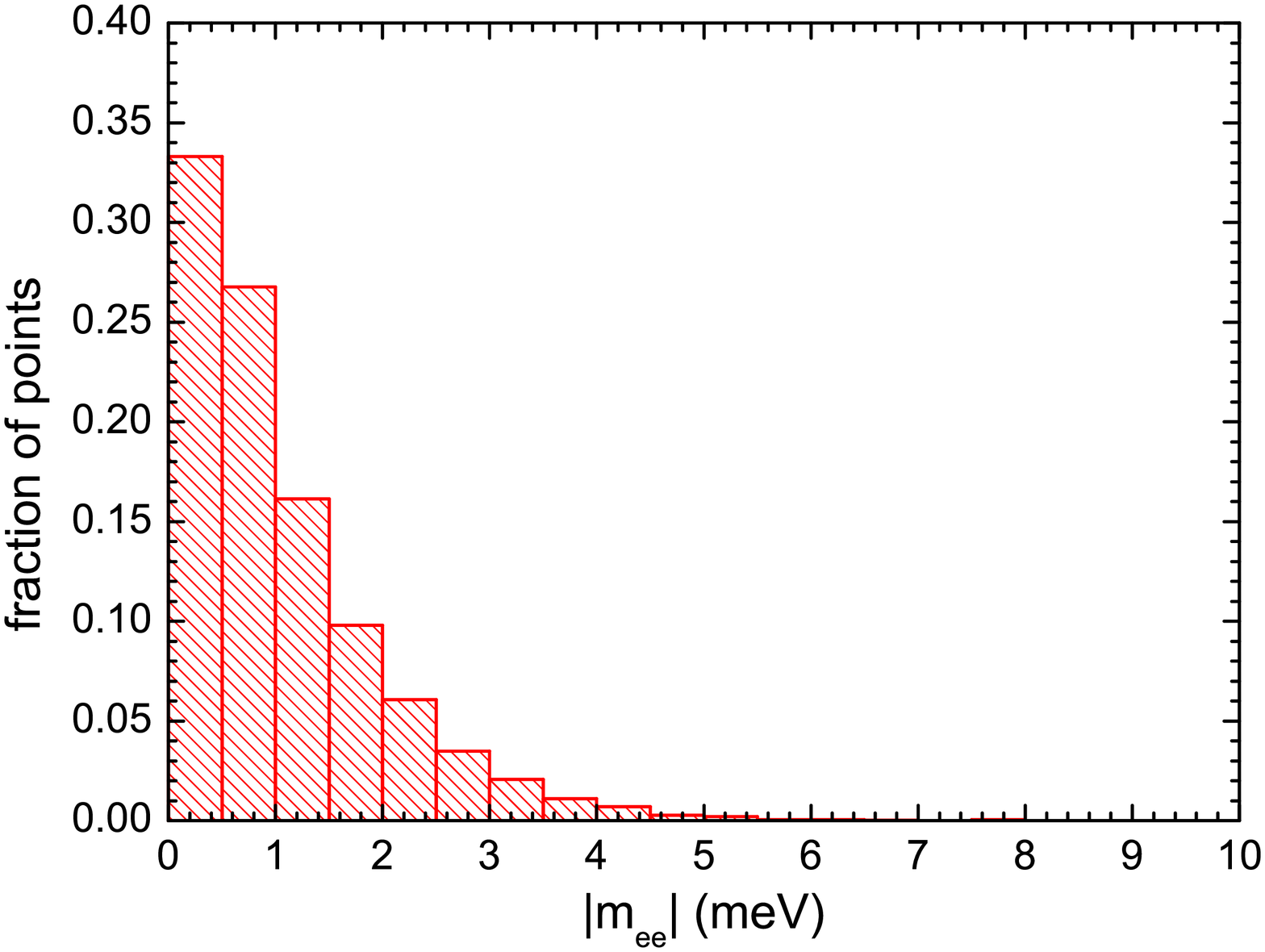}   &  \hskip-1cm   \includegraphics[scale=.30]{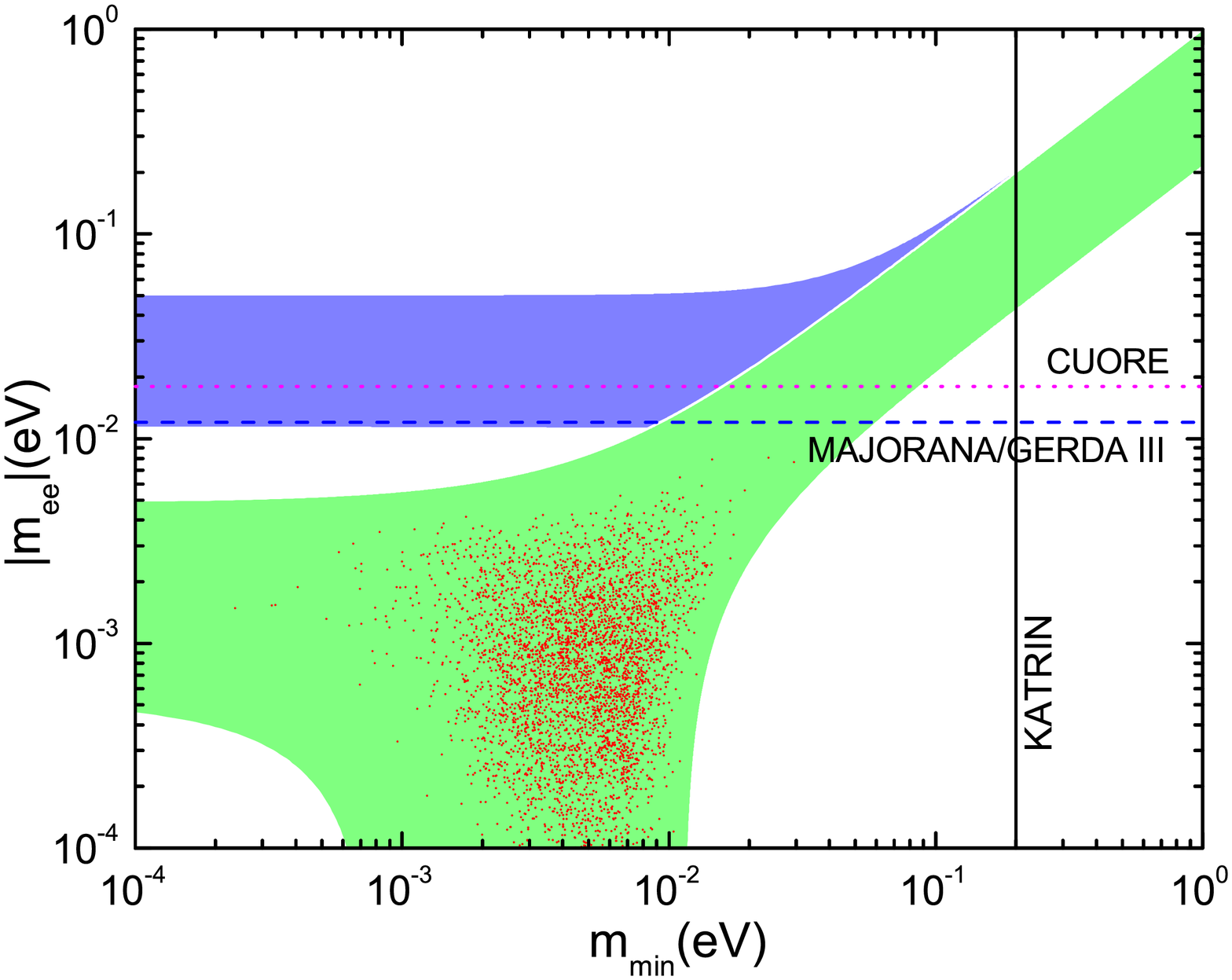}
\end{tabular}
\caption{\label{fig:onu2b_BL1} Histogram of the effective mass
$|m_{ee}|$ (left panel) and the scatter plot of $|m_{ee}|$ versus the lightest neutrino mass $m_{1}$ (right panel) for the $\mathrm{\tt BL_1}$ model. The colored bands represent the regions for the $3\sigma$ ranges of the oscillation parameters in the normal and inverted neutrino mass spectrum respectively. The future sensitivity of 0.2 eV of the KATRIN experiment is shown by the vertical solid line, while the future expected bounds on $|m_{ee}|$ from the CUORE and MAJORANA/GERDA III experiments are represented by
horizontal lines. }
\end{center}
\end{figure}

The rare process, neutrinoless double beta decay ($0\nu2\beta$),
constitutes an important probe for the Majorana nature of neutrino and
lepton number violation~\cite{Schechter:1981bd}, a sizable number of
new experiments are currently running, under construction, or in the
planing phase. The histogram for the distribution of the effective
$0\nu2\beta$-decay mass $|m_{ee}|$ and its correlation with the
lightest neutrino mass $m_{1}$ are given in
Fig. \ref{fig:onu2b_BL1}. We also show the future sensitivity on the
lightest neutrino mass of 0.2 eV from the KATRIN experiment
\cite{katrin}. The horizontal lines represent the sensitivities of the
future $0\nu2\beta$-decay experiments CUORE \cite{cuore} and MAJORANA
\cite{majorana}/GERDA III \cite{gerda}, which are approximately 18 meV
and 12 meV respectively. Clealry the expected effective mass
$|m_{ee}|$ is predicted to be far below the sensitivities of the
planned $0\nu2\beta$ experiments.  The reason for this is the strong
destructive interference amongst the three light neutrinos, as seen
in the right panel. As a result, if $0\nu2\beta$ decay will be
detected in the near future, our construction would be ruled out.

To keep our discussion as generic as possible, we describe the light neutrino masses by the effective higher-dimensional Weinberg operators as shown in Eq.(\ref{eq:U(1)}) and Eq.(\ref{eq:U(1)_extended}), which could come from the so-called type I seesaw mechanism, by integrating out the  right-handed neutrinos. It is interesting to note that $U(1)$ flavor  symmetry models have particularly simple factorization properties  \cite{Rasin:1993kj,Grossman:1995hk} : our various predictions for the light neutrino parameters given above, are independent of the $U(1)$ charge assignments of the right-handed neutrinos. For example, suppose we introduce three right-handed neutrinos transforming under the flavor symmetry $U(1)\times Z_3\times Z_4$ as follows:
\begin{equation}
N^c_1 :~(n_1,0,1),\qquad N^{c}_2 :~ (n_2,0,3),\qquad N^c_3 :~ (n_3,2,2)\,,
\end{equation}
where $n_i$, which are positive integers denoting the $U(1)$ charges
of the heavy Majorana neutrinos. Then one can straightforwardly read
out the Dirac neutrino mass matrix $M_D$ and the Majorana mass matrix
$M_N$ of the right-handed neutrinos,
\begin{equation}
M_D\sim\left(\begin{array}{ccc}
\lambda^{5+n_1}  &  \lambda^{7+n_2}   &   \lambda^{5+n_3}  \\
\lambda^{8+n_1}  &  \lambda^{6+n_2}   &   \lambda^{4+n_3}  \\
 \lambda^{4+n_1}  &  \lambda^{2+n_2}  &  \lambda^{4+n_3}
\end{array}\right)v_u,\qquad M_N\sim\left(\begin{array}{ccc}
\lambda^{2+2n_1}   &  \lambda^{n_1+n_2}   &   \lambda^{5+n_1+n_3}  \\
\lambda^{n_1+n_2}  & \lambda^{2+2n_2}  &   \lambda^{3+n_2+n_3}   \\
\lambda^{5+n_1+n_3}  &  \lambda^{3+n_2+n_3}  & \lambda^{1+2n_3}
\end{array}\right)\Lambda\,.
\end{equation}
The resulting effective light neutrino mass matrix is given by the
seesaw formula
\begin{equation}
M_{\nu}=-M_DM^{-1}_NM^{T}_D\sim\left(\begin{array}{ccc}
\lambda^{9}  &   \lambda^{8}    &    \lambda^{7}   \\
\lambda^{8}  &    \lambda^{7}   &   \lambda^{7}   \\
\lambda^{7}  &    \lambda^{7}   &   \lambda^{6}
\end{array}\right)\frac{v^2_u}{\Lambda}\,.
\end{equation}
This is the same as obtained in the above effective approach given in
Eq.(\ref{eq:nmm}) except that the smallest element $(M_{\nu})_{11}$ is
of order $\lambda^9$ instead of $\lambda^{12}$, both of them are too
small to affect the predictions for the neutrino oscillation
parameters. We get the same light neutrino masses in
Eq.(\ref{eq:mass_BL1}) and the same neutrino mixing angles in
Eq.(\ref{eq:mixing_BL1}) as in the above effective Weinberg operator
neutrino mass generation. We would like to emphasize again that the
predictions for the neutrino masses and mixing parameters are
independent of the charges $n_i$, which drop out in the seesaw formula
for the light neutrino mass matrix. However, different values of the
charges $n_i$ obviously give rise to different Dirac neutrino Yukawa
coupling $Y_{\nu}\equiv M_D/v_u$. As a result, the predictions for
charged lepton flavor violation (LFV) processes such as
$\mu\rightarrow e\gamma$, $\tau\rightarrow\mu\gamma$ and
$\mu\rightarrow3e$ are quite different
\cite{Cannoni:2013gq}. Recalling that the branching ratio of the LFV
process is generally proportional to $Y^4_{\nu}$, the stringent bound
on LFV, in particular from $\mu\rightarrow e\gamma$, can be easily
satisfied for only slightly large $n_i$~\cite{Cannoni:2013gq} while
keeping the predictions for neutrino parameters intact.

\section{\label{sec:BL2}Model for $\mathrm{\tt BL_2}$ mixing }

As explained in section~\ref{sec:th}, the order one atmospheric
neutrino mixing $\sin\theta_{23}\sim1$ generically implies that the
corresponding masses of $\nu_2$ and $\nu_3$ are of the same order of
magnitude within pure $U(1)$ family symmetry schemes. As a result, the
neutrino mass spectrum is quasi-degenerate and strong fine-tuning is
required in order to account for the measured mass-squared differences
$\Delta m^2_{\mathrm{sol}}$ and $\Delta
m^2_{\mathrm{atm}}$~. Furthermore, the renormalization group evolution
effects could drastically enhance the neutrino mixing angles due to
the degeneracy, so that the $\mathrm{\tt BL_2}$ texture would be
spoiled at the electroweak scale.  This can be avoided by extending
the flavor symmetry to $U(1)\times Z_m$. Now the whole flavor symmetry
is chosen to be $U(1)\times Z_2$, the lepton fields carry the
following $U(1)\times Z_2$ charges:
\begin{eqnarray}
\nonumber&&L_{1} :~(3,0),\quad\quad  L_{2} :~(3,1),\quad\quad L_{3} :~(2,0),\quad\quad  \\
&&E^c_1 :~ (4,0),\quad\quad  E^c_2 :~ (2,1),\quad\quad  E^c_3 :~ (0,1),\quad\quad
\end{eqnarray}
Then the light neutrino mass matrix is given, apart from the order
one coefficients, as
\begin{eqnarray}
\label{eq:mass_BL2}&&M_{\nu}\sim\left(
\begin{array}{ccc}
\lambda^{6}  &  \lambda^7  & \lambda^5  \\
\lambda^7  &  \lambda^6  & \lambda^6 \\
\lambda^5  &  \lambda^6  & \lambda^4
\end{array}
\right)\frac{v^2_u}{\Lambda}\,,
\end{eqnarray}
which yields
\begin{eqnarray}
\label{eq:nu_mass}m_{1}\sim\lambda^6\,v^2_u/\Lambda,\quad m_{2}\sim\lambda^6\,v^2_u/\Lambda,\quad m_{3}\sim\lambda^4\,v^2_u/\Lambda
\end{eqnarray}
One sees that the first two light neutrinos are quasi-degenerate in
this model, and their masses are suppressed by
$\mathcal{O}(\lambda^2)$ with respect to the third one.  This
prediction is consistent with the observation that the solar neutrino
mass difference $\Delta m^2_{\mathrm{sol}}$ is much smaller than the
atmospheric neutrino mass difference $\Delta
m^2_{\mathrm{atm}}$. Moreover, the neutrino mass spectrum is predicted to be of the normal hierarchy type here, the same as in the previous $\mathrm{\tt BL_1}$ model (this is also confirmed our numerical analysis). The next generation of higher precision neutrino oscillation experiments is designed to be able to measure neutrino mass hierarchy and the CP phase  \cite{Zhan:2009rs}. Should the latter be determined to be of the inverted type by future experiments, both of our models would be ruled out. On the other hand, the charged lepton mass matrix takes the following form:
\begin{equation}
M_{e}\sim\left(\begin{array}{ccc}
\lambda^{7} & \lambda^6   & \lambda^4 \\
\lambda^{8} & \lambda^5   & \lambda^{3} \\
\lambda^6 &  \lambda^5   & \lambda^3
\end{array}\right)v_d\,,
\end{equation}
which has a ``lopsided'' structure, a large 2-3 mixing arises from the
diagonalization of $M_e$. Obviously it also gives the correct order of
magnitude for the charged lepton mass ratios. Combining the contribution
from both the neutrino and the charged lepton mass matrices diagonalization, the leptonic mixing angles are given by
\begin{equation}
\label{eq:mixing_BL2}\sin\theta_{12}\sim\lambda,\quad\quad\sin\theta_{13}\sim\lambda\,\quad\quad \sin\theta_{23}\sim1\,.
\end{equation}
This is exactly the desired $\mathrm{\tt BL_2}$ mixing pattern,
Eq.~(\ref{eq:BL2}). Here we would like to point out that since the
Super-Kamiokande data indicted large atmospheric neutrino mixing, perhaps
even maximal~\cite{Fukuda:1998mi}, there have been several attempts to
account for the large atmospheric neutrino mixing
$\sin\theta_{23}\sim1$ in terms of Abelian flavor symmetries
\cite{U(1)_lepton}. However, it was usually assumed that the reactor
angle $\theta_{13}$ was rather small, at most of order $\lambda^2$ at
that time~\cite{Maltoni:2004ei}. In contrast, in our construction the
consistency between large $\sin\theta_{23}$ and sizeable
$\sin\theta_{13}$ mixing angles emerges naturally.

\begin{figure}[t!]
\begin{center}
\begin{tabular}{ccc}
\includegraphics[scale=.22]{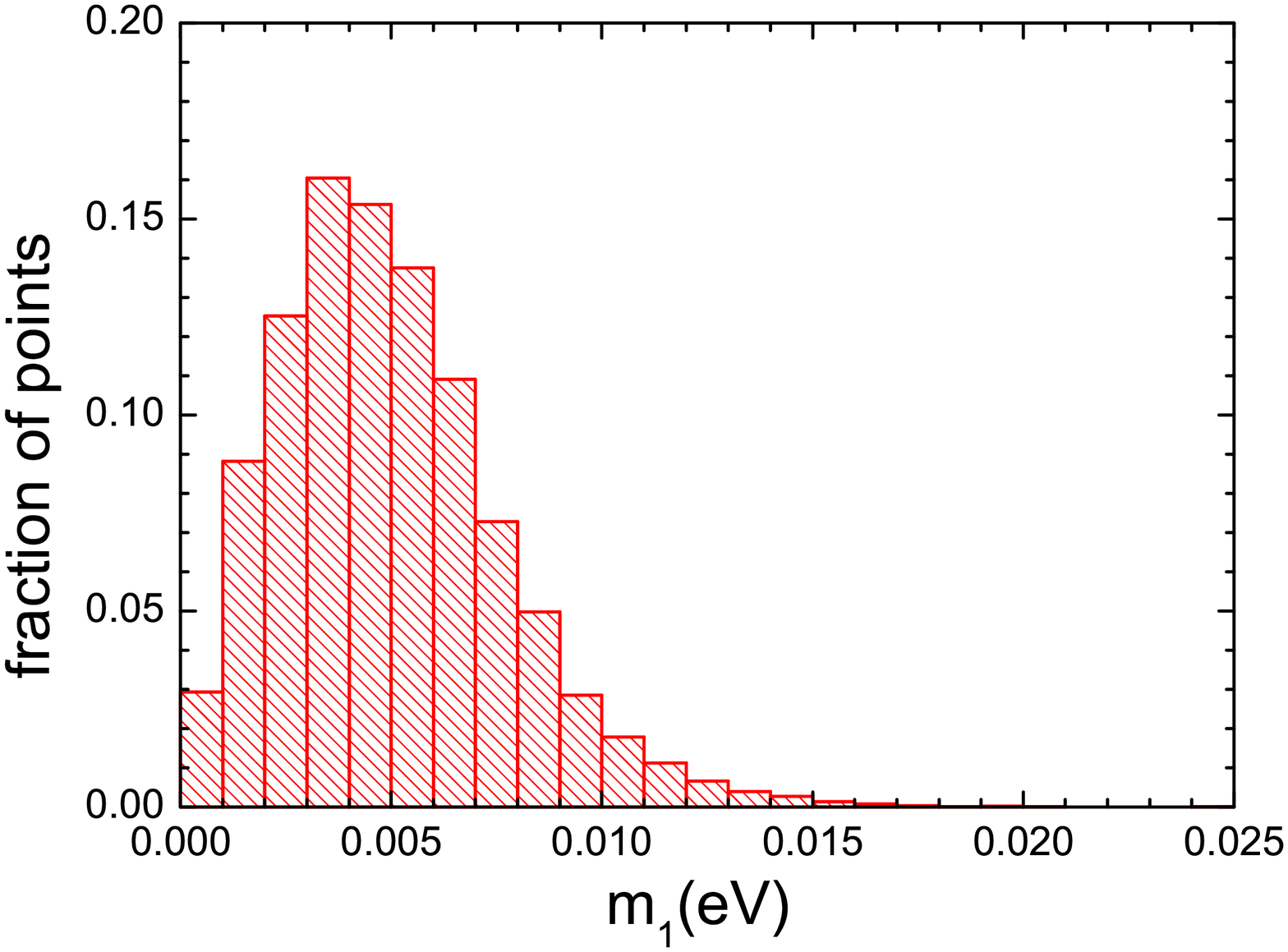}            & \hskip-1cm  \includegraphics[scale=.22]{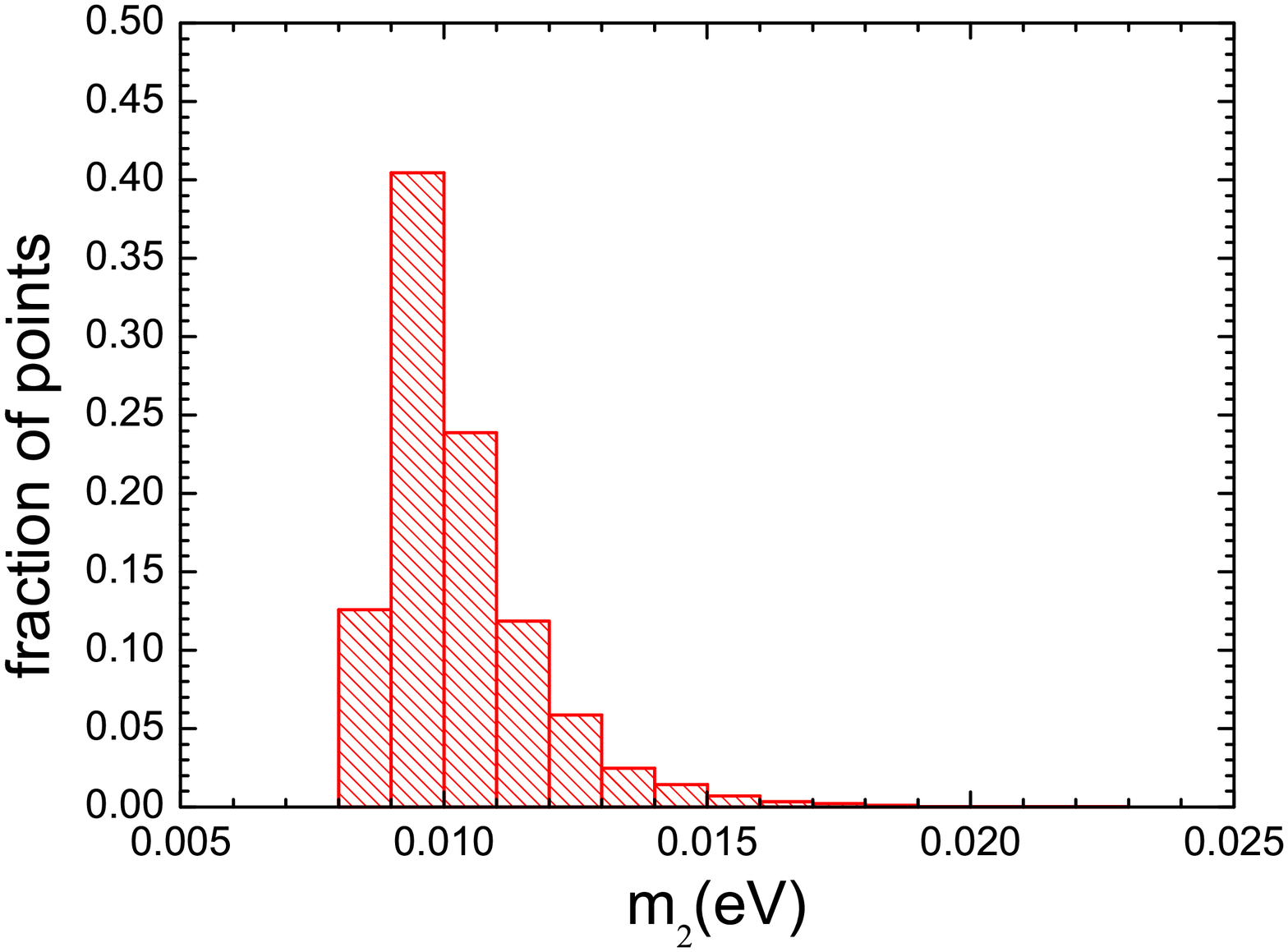}          &  \hskip-1cm   \includegraphics[scale=.22]{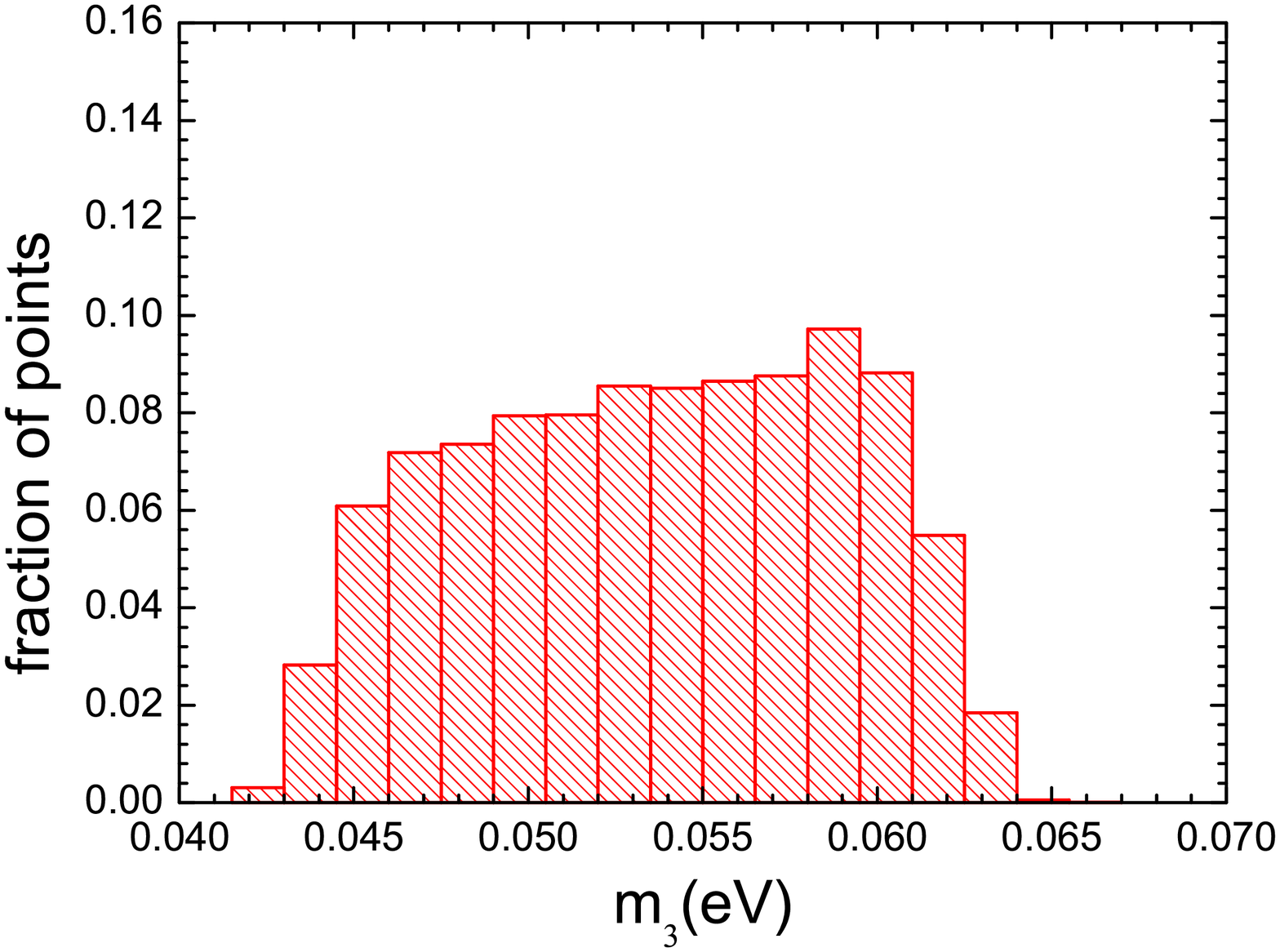} \\
\end{tabular}
\caption{\label{fig:mixing_parameter_BL2} Light neutrino masses in the
  $\mathrm{\tt BL_2}$ model. }
\end{center}
\end{figure}

\begin{figure}[t]
\begin{center}
\begin{tabular}{cc}
\hskip-0.5cm \includegraphics[scale=.30]{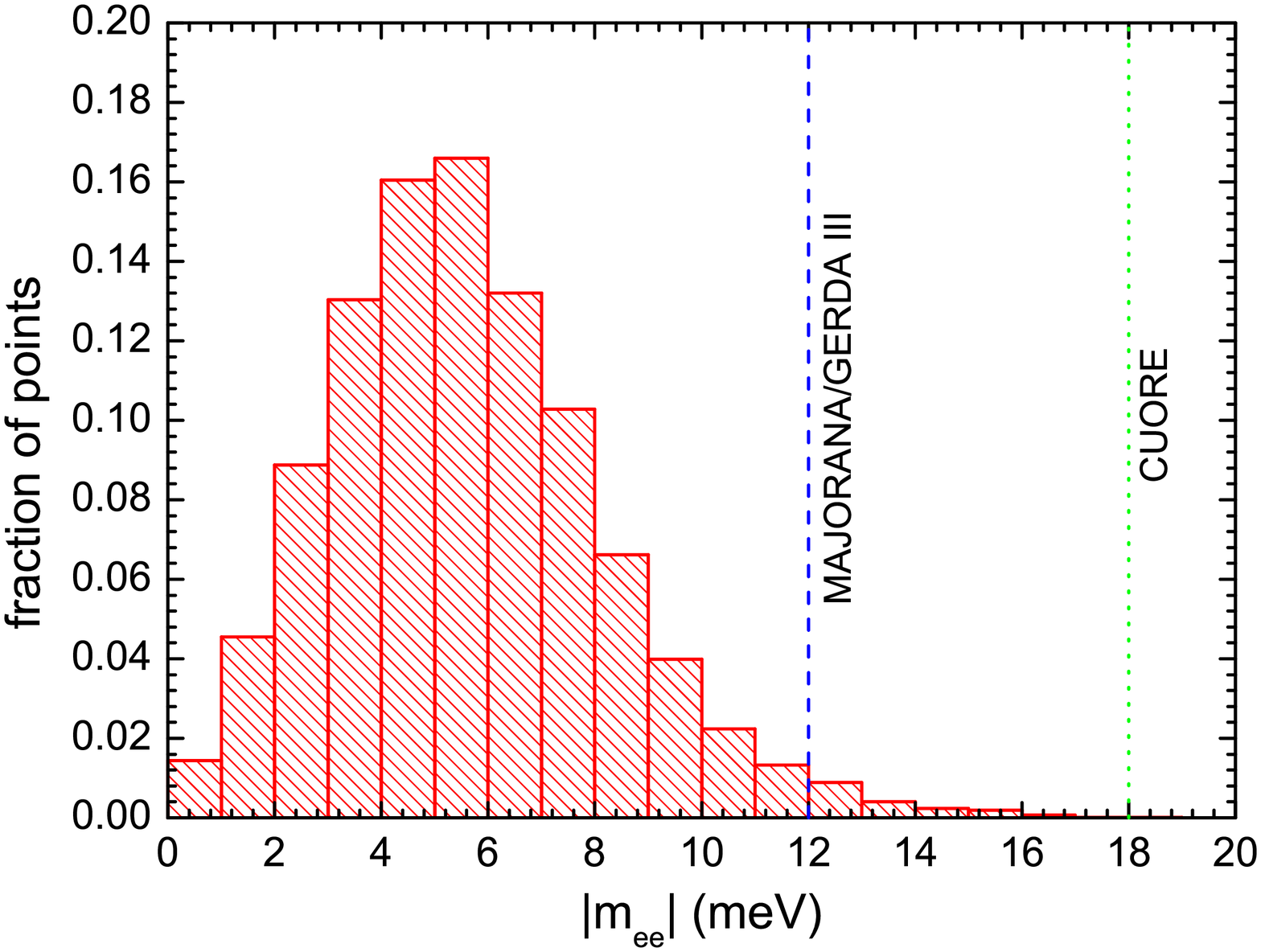}   &  \hskip-1cm   \includegraphics[scale=.30]{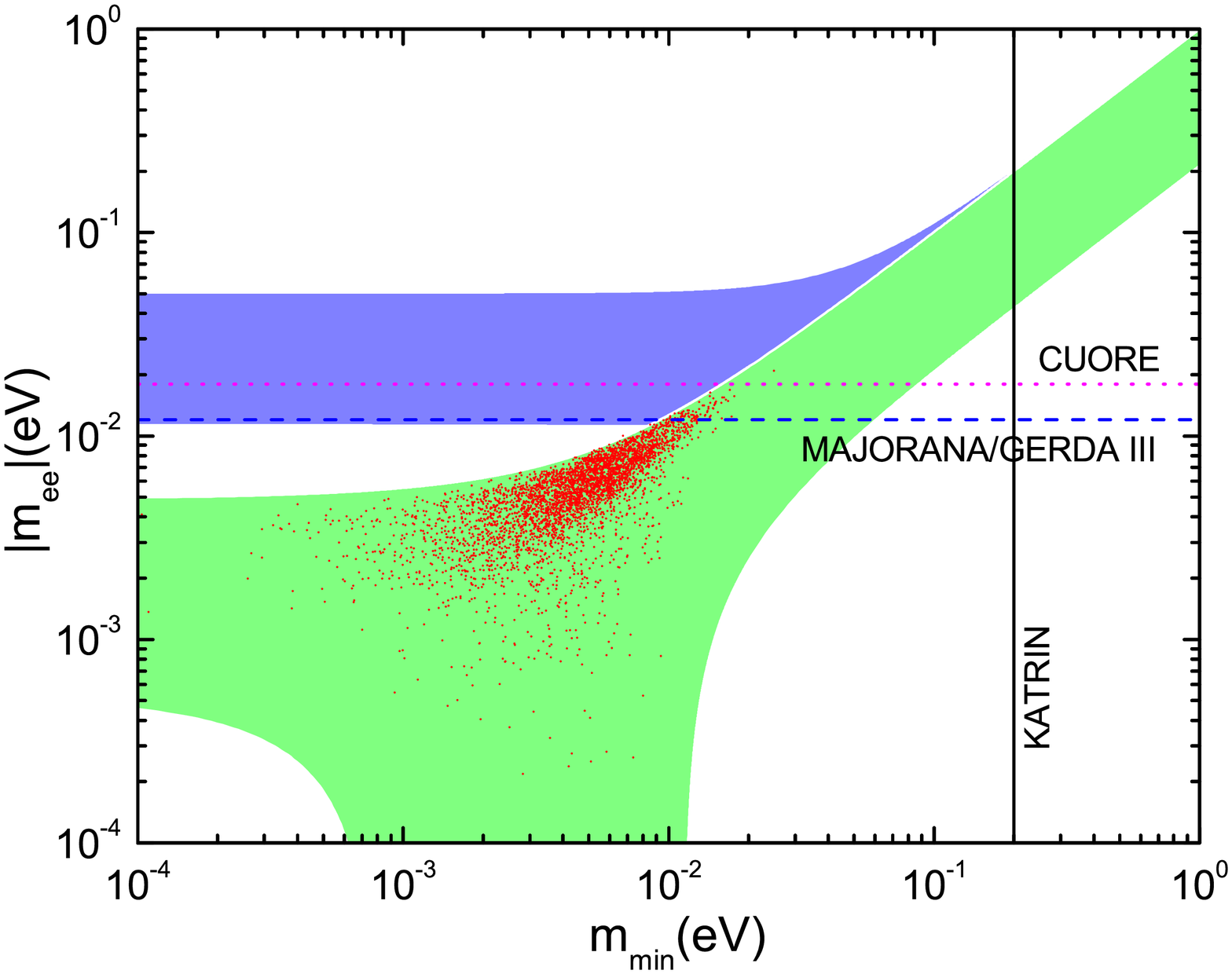}
\end{tabular}
\caption{\label{fig:onu2b_BL2} Histogram of the effective mass
  $|m_{ee}|$ (left panel) and the $|m_{ee}|$ versus the lightest neutrino mass $m_{1}$
  correlation (right panel) predicted in the $\mathrm{\tt BL_2}$
  model. }
\end{center}
\end{figure}

In what follows, we extend the model to include quarks within the
$SU(5)$ unified framework. The fields $Q_i$ and $U^c_i$ together with
$E^c_i$ within the same generation fill out the ${\bf 10}$
representation, while $D^c_i$ and the left-handed lepton doublet $L_i$
make up the $\overline{\mathbf{5}}$ representation. As a result, we
can determine the transformation properties of the quark fields under
the $U(1)\times Z_2$ flavor symmetry as follows:
\begin{eqnarray}
\nonumber&& Q_{1} :~ (4,0),\quad\quad  Q_{2} :~ (2,1),\quad\quad  Q_{3} :~ (0,1),\quad\quad  \\
\nonumber&& U^c_{1} :~ (4,0),\quad\quad\,  U^c_{2} :~ (2,1),\quad\quad  U^c_{3} :~ (0,1),\quad\quad  \\
&&D^c_{1} :~(3,0),\quad\quad\,  D^c_{2} :~(3,1),\quad\quad D^c_{3} :~(2,0)\,. \quad\quad
\end{eqnarray}
The up and down quark mass matrices can be determined in a straightforward way as follows:
\begin{eqnarray}
M_{u}\sim\left(
\begin{array}{ccc}
 \lambda ^8 & \lambda ^7 & \lambda ^5 \\
 \lambda ^7 & \lambda ^4 & \lambda ^2 \\
 \lambda ^5 & \lambda ^2 & 1
\end{array}
\right) v_u,\qquad   M_d\sim\left(
\begin{array}{ccc}
 \lambda ^7 & \lambda ^8 & \lambda ^6 \\
 \lambda ^6 & \lambda ^5 & \lambda ^5 \\
 \lambda ^4 & \lambda ^3 & \lambda ^3
\end{array}
\right) v_d\,.
\end{eqnarray}
which lead to
\begin{eqnarray}
\nonumber &|V_{us}|\sim\lambda,\quad\quad |V_{cb}|\sim\lambda^2,\quad\quad|V_{ub}|\sim\lambda^3\,, \\
&\frac{m_u}{m_c}\sim\lambda^4,\quad\quad \frac{m_c}{m_t}\sim\lambda^4,\quad\quad\frac{m_d}{m_s}\sim\lambda^2,\quad\quad \frac{m_s}{m_b}\sim\lambda^2,\quad\quad \frac{m_b}{m_t}\sim\lambda^3\,,
\end{eqnarray}
which are in excellent agreement with observed quark mass hierarchies
and CKM mixing angles. As in section~\ref{sec:BL1}, we perform a
numerical simulation of the expected neutrino oscillation parameters.
In Fig. \ref{fig:mixing_parameter_BL2} we display the resulting
histograms for the neutrino mass eigenvalues\footnote{Insofar as the
  neutrino mixing angles $\theta_{ij}$ and CP phases $\delta$,
  $\varphi_1$ and $\varphi_2$ are concerned, we do not obtain any
  special predicted pattern, hence the results are not displayed.}.
As expected on the basis of the qualitative estimate in
Eq. (\ref{eq:nu_mass}), the light neutrino mass spectrum is normal
hierarchy, the degenerate spectrum being strongly disfavored, and
almost all the generated points lie in the region of the lightest
neutrino mass $m_{1}$ smaller than 0.015 eV.  The neutrinoless double
beta decay predictions are shown in Fig. \ref{fig:onu2b_BL2}. One sees
that, in contrast with the $\mathrm{\tt BL_1}$ case, although the
effective mass $|m_{ee}|$ is also quite small, with $|m_{ee}|$ around
5 meV preferred, there is a small portion of the parameter space of
the model where the predictions for $|m_{ee}|$ approach the future
experimental sensitivities. However, the points above the sensitivity limits on next generation experiments are statistically rather low. Therefore, if the signal of $0\nu2\beta$ decay would be observed by upcoming experiments, the present $\mathrm{\tt BL_2}$ model would also be ruled out, although not completely. We expect that the future $0\nu2\beta$-decay experiments with sensitivity much higher than MAJORANA/GERDA III should be able to provide a better test of the model.

Now we turn to the seesaw realization of this model, the assignments for the right-handed neutrinos are not unique. As an example, we can introduce three right-handed neutrinos transforming as
\begin{equation}
N^{c}_1 : ~ (n_1,0),\qquad N^{c}_2 : ~ (n_2,1),\qquad N^{c}_3 : ~ (n_3,0)\,.
\end{equation}
Then we obtain the Dirac neutrino mass matrix $M_D$ as well as the
right-handed neutrino mass matrix $M_N$,
\begin{equation}
M_D\sim\left(\begin{array}{ccc}
\lambda^{3+n_1}   & \lambda^{4+n_2}  & \lambda^{3+n_3}  \\
\lambda^{4+n_1}   & \lambda^{3+n_2}  &  \lambda^{4+n_3}  \\
\lambda^{2+n_1}   & \lambda^{3+n_2}  &  \lambda^{2+n_3}
\end{array}\right)v_u,\quad M_N\sim\left(\begin{array}{ccc}
\lambda^{2n_1}    &   \lambda^{1+n_1+n_2}   &  \lambda^{n_1+n_3}  \\
\lambda^{1+n_1+n_2}   &   \lambda^{2n_2}    &  \lambda^{1+n_2+n_3}  \\
\lambda^{n_1+n_3}    &  \lambda^{1+n_2+n_3}  & \lambda^{2n_3}
\end{array}\right)\Lambda\,.
\end{equation}
The effective light neutrino mass matrix is
given by the seesaw relation
\begin{equation}
M_{\nu}=-M_DM^{-1}_MM^{T}_D\sim\left(\begin{array}{ccc}
\lambda^{6}  &   \lambda^{7}   &  \lambda^{5}   \\
\lambda^{7}  &   \lambda^{6}   &  \lambda^{6}  \\
\lambda^{5}  &   \lambda^{6}   &  \lambda^{4}
\end{array}\right)\frac{v^2_u}{\Lambda}
\end{equation}
This is exactly Eq.(\ref{eq:mass_BL2}), consequently the predictions
for neutrino parameters in Eq.(\ref{eq:nu_mass}) and
Eq.(\ref{eq:mixing_BL2}) remain, note that dependence on the
right-handed neutrino charges $n_i$ drops out. However, different
values of the charges $n_i$ result in different LFV predictions, and
the model would be less constrained for slightly large $n_i$
assignments~\cite{Cannoni:2013gq}.

\section{\label{sec:conclusion}Conclusions}

The recent neutrino oscillation experimental highlights: (i) rather
large value of reactor mixing angle $\theta_{13}$ and (ii) indication
of significant deviation of the atmospheric neutrino mixing angle
$\theta_{23}$ from maximality may change our theoretical approach for
constructing neutrino mass models. In this paper, we study the
Wolfenstein-like mixing schemes: $\mathrm{\tt BL_1}$ mixing in which
$\sin\theta_{12}\sim\lambda$, $\sin\theta_{13}\sim\lambda$,
$\sin\theta_{23}\sim\lambda$, and $\mathrm{\tt BL_2}$ mixing, in which
$\sin\theta_{12}\sim\lambda$, $\sin\theta_{13}\sim\lambda$,
$\sin\theta_{23}\sim 1$. The largish $\theta_{13}$ can be naturally
accommodated in both of them, the two mixing patterns differ in the
order of magnitude of $\sin\theta_{23}$, the $\mathrm{\tt BL_1}$
texture is favored for $\theta_{23}$ in the first octant, while
$\mathrm{\tt BL_2}$ is preferred for the second octant
$\theta_{23}$. In order to produce the $\mathrm{\tt BL_1}$ mixing
without invoking unnatural cancellation, the Abelian flavor symmetry
should be $U(1)\times Z_m\times Z_n$ with the parity of $m$ and $n$
being opposite. A concrete model based on $U(1)\times Z_3\times Z_4$
family symmetry is constructed, where the light neutrino mass
hierarchy $m_{2}/m_{3}\sim\lambda$ is realized due to the
discrete nature of $Z_3\times Z_4$. The ratio $\Delta
m^2_{\mathrm{sol}}/\Delta m^2_{\mathrm{atm}}$ is expected to be of
order $\lambda^2$ in this model, which is in good agreement with
experimental data in contrast with conventional $U(1)$ or $U(1)\times
Z_m$ flavor symmetry constructions.
Furthermore, the model is embedded into the $SU(5)$ grand unified
theory to describe the quark masses and mixing simultaneously. As for
the $\mathrm{\tt BL_2}$ mixing, it can be reproduced within the
framework of pure $U(1)$ flavor symmetry. However, the light neutrino
mass spectrum is expected to be quasi-degenerate, hence fine-tuning of
the neutrino mass parameters is needed in order to achieve the
observed mass-squared differences. To improve upon this situation, the
family symmetry is enlarged to $U(1)\times Z_2$, which gives rise to
both large atmospheric neutrino mixing $\sin\theta_{23}\sim1$ and
hierarchical neutrino masses. The model is extended to $SU(5)$ grand
unified theory as well.

We show that both models can give a successful description of the
observed quark and lepton masses and mixing angles, and the numerical
results are nicely in agreement with the theoretical estimates and the
qualitative discussions. The light neutrinos are normal mass hierarchy
in both models, quasi-degenerate spectrum is strongly disfavored. If the next generation high precision neutrino oscillation experiments determine that the neutrino mass spectrum is inverted hierarchy, both our constructions will be ruled out. The present framework can not predict the CP violating phases $\delta$, $\varphi_1$ and $\varphi_2$. The $0\nu2\beta$-decay effective mass $|m_{ee}|$ is predicted to be rather small in both constructions, substantial part of the data are below the sensitivity of future experiments except for a region of the $\mathrm{\tt BL_2}$ model indicated in Fig.~\ref{fig:onu2b_BL2}.  Therefore future
$0\nu2\beta$-decay experiments such as CUORE, MAJORANA and GERDA III
will provide another important test of the present models.%

\section*{Acknowledgements}

This work was supported by the National Natural Science Foundation of
China under Grant No~10905053, Chinese Academy KJCX2-YW-N29, DFG grant
WI 2639/4-1 and the 973 project with Grant No.~2009CB825200; by the
Spanish MINECO under grants FPA2011-22975 and MULTIDARK CSD2009-00064
(Consolider-Ingenio 2010 Programme), by Prometeo/2009/091 (Generalitat
Valenciana), and by the EU ITN UNILHC PITN-GA-2009-237920.  S.M. is
supported by a Juan de la Cierva grant.

\renewcommand{\baselinestretch}{1}%

\end{document}